\input cp-aa.tex

%
\def\logt{\ifmmode{\log T_{\rm eff}}\else{$\log T_{\rm eff}$}\fi}
\def\logg{\ifmmode{\log g}\else{$\log g$}\fi}
\def\teff{\ifmmode{T_{\rm eff}}\else{$T_{\rm eff}$}\fi}
\def\logl{\ifmmode{\log L/{L_\odot}}\else{$\log L/{L_\odot}$}\fi}
\def\mdot{\ifmmode{\dot M}\else{$\dot M$}\fi}
\def\vterm{\ifmmode{v_\infty}\else{$v_\infty$}\fi}
\def\bminu{$[B-U]$}
\def\bminl{$[B-L]$}
\def\kms{\ifmmode{~{\rm km\,s}^{-1}}\else{~km\thinspace s$^{-1}$}\fi}
\def\Kkms{\ifmmode{~{\rm K\, km\, s}^{-1}}\else{K\thinspace km\thinspace
                         s$^{-1}$}\fi}
\def\ergs{\ifmmode{~{\rm ergs}}\else{~ergs}\fi}
\def\ergps{\ifmmode{~{\rm ergs\,s}^{-1}}\else{~ergs\thinspace
                          s$^{-1}$}\fi}
\def\pc{\ifmmode{~{\rm pc}}\else{~pc}\fi}
\def\kpc{\ifmmode{~{\rm kpc}}\else{~kpc}\fi}
\def\cm2{\ifmmode{~{\rm cm}^{-2}}\else{~cm$^{-2}$}\fi}
\def\cc{\ifmmode{~{\rm cm}^{-3}}\else{~cm$^{-3}$}\fi}
\def\yr{\ifmmode{~{\rm yr}}\else{~yr}\fi}
\def\Myr{\ifmmode{~{\rm Myr}}\else{~Myr}\fi}
\def\magn{\ifmmode{^{\rm m}}\else{$^{\rm m}$}\fi}
\def\um{\ifmmode{~\mu {\rm m}}\else{~$\mu {\rm m}$}\fi}
\def\mas{\ifmmode{~{\rm mas}\,{\rm yr}^{-1}}\else{~mas\thinspace
                                                      yr$^{-1}$}\fi}
\def\arcsec{\ifmmode{^{\prime\prime}}\else{$^{\prime\prime}$}\fi}
\def\farcs{\ifmmode{.\kern-0.3em^{\prime\prime}}\else
             {.\kern-0.3em$^{\prime\prime}$}\fi}
\def\av{\ifmmode{A_V}\else{$A_V$}\fi}
\def\HI{H\kern .125em {\sc i}}
\def\HII{H\kern .125em {\sc ii}}
\def\nh{N(H\kern .125em {\sc i})}

\def\ionI#1#2{#1\kern .125em {\sc i}~$\lambda#2$}
\def\ionII#1#2{#1\kern .125em {\sc ii}~$\lambda#2$}
\def\vsini{\ifmmode{v\kern-.1em\sin\kern-.1em i}
            \else{$v\kern-.1em\sin\kern-.1em i$}\fi}
\def\vbreak{\ifmmode{v_{\rm br}}\else{$v_{\rm br}$}\fi}

\def\scrscr{\scriptscriptstyle}
\def\veq{\ifmmode{v_{\rm e}}\else{$v_{\rm e}$}\fi}
\def\msun{\ifmmode{{\rm\ M}_\odot}\else{${\rm\ M}_\odot$}\fi}
\def\lessapprox{\,\raise 0.6ex\hbox{$<$}\kern -0.75em\lower 0.47ex
    \hbox{$\sim$}\,}
\def\largapprox{\,\raise 0.6ex\hbox{$>$}\kern -0.75em\lower 0.47ex
    \hbox{$\sim$}\,}
\def\deg#1{\ifmmode{#1{^\circ}}\else{$#1{^\circ}$}\fi}
\def\degpnt{^{\circ} \kern -0.4em \hbox{.}\,}
\def\magndot{\ifmmode{\kern.1em.\kern-0.4em^{\rm
     m}}\else{\kern.1em.\kern-0.4em$^{\rm m}$}\fi} 

%
\def\runin#1#2{\smallskip\noindent#1\ #2}
 
%
%
%
 
%
\def\.#1{{\accent"C7 #1}}
\def\dot{\mathaccent"70C7 }
\newdimen\aadimen
\def\AA{\leavevmode\setbox0\hbox{h}\aadimen\ht0
  \advance\aadimen-1ex\setbox0\hbox{A}\rlap{\raise.67\aadimen
  \hbox to \wd0{\hss\char'27\hss}}A}

%
%
\def\ea{{\rm et~al.}}
\def\aj#1{AJ\ #1}
\def\apj#1{ApJ\ #1}
\def\aap#1{A\&A\ #1}
\def\araa#1{ARA\&A\ #1}
\def\aapss#1{A\&AS\ #1}
\def\apjss#1{ApJS\ #1}
\def\apss#1{Ap\&SS\ #1}
\def\mn#1{MNRAS\ #1}
\def\pasp#1{PASP\ #1}

%
%
\input psfig

%
%
  \MAINTITLE={ High S/N Echelle Spectroscopy in Young Stellar Groups
               \FOOTNOTE{ Based on observations obtained at the
               European Southern Observatory (ESO), La Silla, Chile in
               the framework of Key Programme 5-005-45K} }

  \SUBTITLE={ II. Rotational Velocities of Early-Type Stars in Sco OB2 }
  \AUTHOR={ A.G.A.\ Brown@1, W.\ Verschueren@{2,1} }
  \INSTITUTE={ @1 Sterrewacht Leiden, P.O.~Box 9513, 2300 RA, Leiden,
		  The Netherlands 
               @2 University of Antwerp (RUCA), Astrophysics Research
                  Group, Groenenborgerlaan 171, 2020 Antwerpen, Belgium }
  \ABSTRACT={ We investigate the rotational velocities of early-type
stars in the Sco OB2 association. We measure \vsini\ for 156
established and probable members of the association. The measurements
are performed with three different techniques, which are in increasing
order of expected \vsini: 1) converting the widths of spectral lines
directly to \vsini, 2) comparing artificially broadened spectra of low
\vsini\ stars to the target spectrum, 3) comparing the \ionI{He}4026\
line profile to theoretical models. The sample is extended with
literature data for 47 established members of Sco OB2. Analysis of the
\vsini\ distributions shows that there are no significant differences
between the subgroups of Sco OB2. We find that members of the binary
population of Sco OB2 on the whole rotate more slowly than the single
stars. In addition, we find that the B7--B9 single star members rotate
significantly faster than their B0--B6 counterparts. We test various
hypotheses for the distribution of \vsini\ in the association. The
results show that we cannot clearly exclude any form of random
distribution of the direction and/or magnitude of the intrinsic
rotational velocity vector. We also investigate the effects of
rotation on colours in the Walraven photometric system. We show that
positions of B7--B9 single dwarfs above the main sequence are a
consequence of rotation. This establishes the influence of rotation on
the Walraven colours, due primarily to surface gravity effects. }
  \KEYWORDS={ stars: early-type; formation; rotation -- open clusters
              and associations: individual: Sco OB2 }
  \THESAURUS={ 08.05.1; 08.06.2; 08.18.1; 10.15.2 Sco OB2 }
  \OFFPRINTS={ \hfil\break A.G.A.~Brown, brown\at strw.leidenuniv.nl }
  \DATE={ Received\dots, accepted\dots }
\maketitle

\MAINTITLERUNNINGHEAD{ Rotational Velocities of Early-Type Stars in
Sco OB2 }

\AUTHORRUNNINGHEAD{ A.G.A.~Brown \& W.~Verschueren }

\titlea{Introduction}
An outstanding problem in theories of star formation is the so-called
angular momentum problem. Molecular clouds and cloud cores, even if
they spin about their axis only once per Galactic rotation, contain
much more angular momentum than the stars that form from them (see
e.g., Spitzer 1968). Observations suggest that an interstellar parent
cloud must lose at least two to four orders of magnitude of angular
momentum for a relatively wide binary system and a single star,
respectively, to become dynamically possible (see Mouschovias
1991). How does angular momentum get redistributed when stars form and
what is the resulting distribution of rotational velocities? These
questions can best be addressed by studying young stellar groups in
which the stars are unevolved and physically related. It is in these
groups that the observed distribution of projected rotational
velocities (\vsini) may still reflect the distribution of rotational
velocities at the time of star formation. Observations of the present
\vsini\ distribution provide constraints on the angular
momentum history of forming stars, and contain information on
mechanisms that lead to angular momentum redistribution during star
formation.

Suggested mechanisms for the redistribution of angular momentum,
include magnetic braking and the formation of disks. Magnetic braking
is important during the early, diffuse stages of star formation
(Mouschovias 1991). Stellar disks are important during the
proto-stellar and pre main-sequence phases (see e.g., Bodenheimer \ea\
1993). Another way of redistributing angular momentum during star
formation is through the formation of binary systems. This mechanism
can be studied by relating the characteristics of the binary
population in stellar groups to the
\vsini\ distribution, which may also provide information about tidal
interactions in close binaries. Ultimately, observations of the
\vsini\ distribution in young stellar groups may provide information
on the star formation process itself. For a recent review on the role
of rotation in star formation, see Bodenheimer \ea (1993).

Another important aspect of the study of rotational velocities of
stars is the effect of rotation on observed stellar parameters. It is
well known that photometry and spectral classification of stars are
affected by rotation, due to surface gravity effects. The resulting
misinterpretation of the observed stellar parameters will bias age
determinations for stellar groups and the derived mass distributions
(e.g., Maeder 1971).

OB associations are young and mostly unobscured sites of recent star
formation and as such are well suited for studies of the properties of
young groups of stars (for reviews see Blaauw 1964, 1991). In this
paper we focus on the Sco OB2 association. Because of its proximity
(the distance to Sco OB2 is $\sim 145\pc$, see de Geus \ea\ 1989) it
is easily accessible to proper motion studies and as a result
extensive membership determinations have been carried out for this
association. Sco OB2 consists of three well-known subgroups; Upper
Scorpius (US), Upper Centaurus Lupus (UCL) and Lower Centaurus Crux
(LCC). The Upper Scorpius subgroup is located near the remnants of its
parental molecular cloud, and some of its stars are still partly
obscured by gas and dust. We also study a subgroup located southeast
of US, although the physical reality of this subgroup has never been
established.

The ages of the three main subgroups were determined most recently
from Walraven photometry by de Geus \ea (1989). They found ages of
4--$5\Myr$, 14--$15\Myr$ and 11--$12\Myr$ for US, UCL and LCC,
respectively. For US the kinematic age (i.e., the age determined by
tracing the proper motions of the members back in time) was determined
by Blaauw (1978, 1991) to be $5\Myr$.

Studies of rotational velocities in Sco OB2 have in the past
concentrated mainly on US. Slettebak (1968) determined \vsini\ for 82
stars in US and UCL and found that on average the stars in US rotate
faster than those in UCL (174\kms\ vs.\ 119\kms, but the rms spreads
on these numbers are $\sim 100\kms$) and that the mean \vsini\ for
B7--A0 stars is larger than that of corresponding field stars
(212\kms\ vs.\ 138\kms). The B0--B6 stars were found to rotate
somewhat more slowly than field stars. Rajamohan (1976) derived
rotational velocities for 112 members of Sco OB2 and found that for
stars with $M_V<0\magndot0$ the distributions of rotational velocities
are similar for US and UCL and resemble those of field stars. Stars
with $M_V>0\magndot0$, all of which are located in US, were found to
rotate much faster than corresponding field stars.

In this paper we present new projected rotational velocities from
high-resolution spectra for stars that are established or probable
members of the Sco OB2 association. We briefly describe the
observations and sample selection in Sect.~2. In Sect.~3 we
present and discuss the different techniques used to derive \vsini\
from the observed spectra. In Sect.~4 we discuss the observed
distribution of projected rotational velocities. Various hypotheses
about the true distribution of rotational velocities are tested for
the known members of Sco OB2. We also discuss the interpretation of
the data in the context of these hypotheses. In Sect.~5 we combine
photometric data with our rotational velocities and confirm that
rotation affects the colours of the stars in the Walraven photometric
system. We discuss the effects of the colour changes on age
determinations and derivations of mass distributions. Finally, in
Sect.~6 we summarize our conclusions and suggest future work.

\titlea{Observations}
The spectra from which the projected rotational velocities are derived
were all obtained with the ECHELle + Electronographic Camera (ECHELEC)
spectrograph, mounted at the Coud\'e focus of the ESO $1.52$m
telescope at La Silla. It consists of a 31.6 grooves/mm echelle
grating and a 632 grooves/mm grism as a cross-disperser. The detector
was a CCD (thinned, back-illuminated, RCA) with $640\times1024$ pixels
of $15\times15\um^{2}$, and a read-out noise of 65 e$^{-}$. In order
to increase the signal to noise on read-out the CCD pixels were binned
$2\times2$ (henceforth any reference to CCD-pixels is to the binned
pixels). The linear dispersion of ECHELEC is $3.1$~\AA/mm at 4000~\AA\
as measured for our data. This implies a velocity scale of $7\kms$ per
binned pixel and a resolving power of $21\,500$. The spectra cover the
wavelength region 3800--4070~\AA\ which was recorded in 11--12
spectral orders. The formal (i.e., taking only photon and read-out
noise into account) signal to noise ratio, measured at the top of the
blaze profile in each order, typically varies from $\sim 70$ to $\sim
300$ between the blue and the red end of the spectrum. Further details
on the observations are given in Verschueren
\ea (1996), which also describes the data reduction procedure in
detail.

The data were collected during 1991--1993 in the framework of an ESO
Key Programme aimed at the determination of high-precision radial
velocities of early-type stars in young stellar groups (Hensberge
\ea 1990). The stars that were observed are established or probable
members of Sco OB2. They form a subset of a larger sample of stars
that was submitted for observation by the HIPPARCOS satellite. The
details of the selection of the HIPPARCOS sample can be found in de
Zeeuw, Brown \& Verschueren (1994). For the present study a total of
156 stars was observed, of which 136 already have a measured
\vsini. We note here that the observations were not specifically
designed for rotational velocity studies. The aim was to gather radial
velocities and also to observe stars repeatedly in order to detect
possible spectroscopic binaries. However, the data are also well
suited for deriving a large and homogeneous set of rotational
velocities. Because of telescope and instrumental limitations the
magnitude limit of our sample is $\sim 7\magndot5$. As a result 75\%
of the stars in our sample are of spectral type earlier than B7 (see
also Verschueren \ea 1996).

Of the observed sample 74 stars are established members, inferred from
proper motions, found by Blaauw (1946) and Bertiau (1958). For 13
stars membership was inferred from photometry by de Geus
\ea (1989). For the analysis in Sects.~4 and 5 we also added
\vsini\ data from the literature (Slettebak 1968, and Uesugi \&
Fukuda 1981) for 47 established members which were not observed in our
programme. Of these additional stars, 36 are proper motion members,
the remaining ones are photometric members. Our observed sample is
divided among spectral type as follows: 2 O stars, 93 B0--B3 stars, 22
B4--B6 stars, 25 B7--B9 stars and 14 A--F stars. The additional data
from the literature consists of 6 B0--B3 stars, 8 B4--B6 stars, 32
B7--B9 stars and 1 A star. In addition to stars in Sco OB2 8
early-type stars were observed which are not related to the
association. A number of these stars were used in the process of
deriving \vsini\ (see Sect.~3.2). The data were collected during
four observing runs. The runs are labeled E2, E3, E5 and E7 by
Verschueren \ea (1996); we will use these labels here as well.

\titlea{Techniques Used to Determine \vsini}
We have used three different methods to obtain \vsini\ from the
ECHELEC spectra for the stars in our sample:

\medskip
\item{1.} Deriving \vsini\ from the full-width half-maximum \allowbreak
(FWHM) of weak metal lines. This method is applicable for
$\vsini\la\allowbreak 80\kms$.

\item{2.} Deriving \vsini\ from the comparison between an artificially
broadened sharp-lined template spectrum and the spectrum of a star for
which \vsini\ is to be determined. This method is potentially
applicable for $50\kms \la \vsini \la 200\kms$. This range depends on
spectral type (see Sect.~3.2 and Appendix C).

\item{3.} Deriving \vsini\ from the comparison of theoretical line
profiles for the \ionI{He}4026\ line with the observed line profile.
The theoretical profile is taken from atmosphere models for rotating
early-type stars by Collins \ea\ (1991, hereafter CTC). This method is
in principle applicable for all velocities but is most useful for
velocities for which method 2 fails, specifically for higher \vsini.
\medskip

Methods 1 and 2 rely on the classical model for rotating stars, as
presented e.g., in Chapt. 2 of Tassoul (1978). The third method
relies on theoretical models in which the line profiles are calculated
{\it ab initio}. Recently Collins \& Truax (1995) discussed the
consequences of using the classical model rather than the physically
more appropriate ab initio model. The most important point Collins \&
Truax make is to beware of the assumptions that underlie the classical
model for rotating stars. These are:

\medskip
\item{A.} The observational aspect of a uniformly rotating star may be
approximated by a circular disk subject to a limb-darkening law
applicable to all parts of the stellar disk.

\item{B.} The limb-darkening law and coefficient appropriate for the
radiation in a line are the same as for the continuum.

\item{C.} The form of a line does not change over the apparent disk;
it is simply uniformly Doppler-shifted by the radial motion resulting
from rotation.
\medskip

Assumption A means that the radial velocity is constant along lines
parallel to the meridian plane of the star. This assumption breaks
down at high rotational velocities where the surface of the star
starts to deform and can no longer be described as a sphere. The
effects of the deformation of the star are twofold. The lines of
constant radial velocity are not parallel to the meridian plane of the
star and the line formation process is influenced.

The second assumption, B, is flawed for the following reason. The
limb-darkening law describes the variation of the specific emergent
intensity from the centre of the stellar disk to the limb. The
limb-darkening laws that are usually quoted in the literature are
derived for the continuum. However, because the core of a line has a
lower specific intensity at the centre of the stellar disk, the total
variation from centre to limb of the intensity at the wavelength of
the line-core will be less. Hence the limb-darkening coefficient will
be smaller for the line-core. This effect is very important in strong
lines, where the limb-darkening coefficient shows a variation from the
continuum through the wings to the core.

Assumption C does not hold at high rotational velocities because the
distortion of the star introduces variations in the
ionization-excitation equilibrium over the surface of the star.
Furthermore, the local form of the line is influenced by gravity
darkening and shape distortion. This means that there is a difference
between the line profiles of an intrinsically slow rotator and those
of a fast rotator seen pole-on. However, according to calculations in
Collins \& Truax (1995) the differences are very small and of no
concern to the determination of \vsini.

Before discussing the three methods in detail we should mention the
format of the spectra. After the data reduction process the spectra
are in the form of normalized flux as a function of CCD-pixel. The
pixel scale has to be converted to a wavelength scale and for the
radial velocity studies a log-wavelength scale is used (see
Verschueren \ea 1996). It turns out that due to the specific geometry
of the ECHELEC instrument, the CCD-pixel spacing in the spectra is
almost identical to a spacing in $\ln\lambda$.  For methods 1 and 2 we
used spectra that were not yet converted to $\ln\lambda$-space and for
which the different echelle orders were not merged. Furthermore, the
continuum of the spectra was not yet globally corrected for
instrumental effects (such as the echelle blaze), which means we had
to use a local continuum for method 1. For method 3 we only used the
order containing the \ionI{He}4026\ line, and to facilitate the
comparison of the data to the model profiles the order was rebinned
into $\ln\lambda$-space. In the rebinned spectra the spacing is
$2.8\times 10^{-5}$ in $\ln\lambda$ or $8.4\kms$.

\titleb{Deriving \vsini\ for Sharp-Lined Stars}
If one assumes that spectral lines in a star are broadened by rotation
only, the width of the lines can be used to derive \vsini. In practice
spectral lines are also broadened by thermal and instrumental effects
and by other mechanisms, such as the Stark effect (see e.g., Gray
1992, Chapt. 17). Therefore one should use weak metal lines that are
in principle only affected by thermal and instrumental broadening; two
effects one can correct for. However, the weak metal lines can also be
broadened due to blending, the possible spectroscopic binary nature of
the star, or curve of growth effects. The latter is important for some
of the stars in our sample (see below). Possible spectroscopic
binaries show obvious double lines or they reveal themselves as stars
for which \vsini\ changes from one observing run to the
other. Neglecting the binary nature leads to overestimates of
\vsini. The presence of line blends is more or less automatically
taken care of in the conversion of line-widths to \vsini\ described
below.

At low values of \vsini\ one has to take into account the consequences
of assumptions B and C. Because the lines we use are intrinsically
narrow and weak (the median residual flux in the line cores is $\sim
94$\%, the average and rms spread being $\sim 91$\% and $\sim 8$\%) the
limb-darkening parameter will not be very different from the continuum
limb-darkening parameter. The small wavelength range of the ECHELEC
spectra also ensures that the limb-darkening parameter is nearly
constant over the entire spectrum.

\begfig 8cm
\vskip -8cm
\psfig{figure=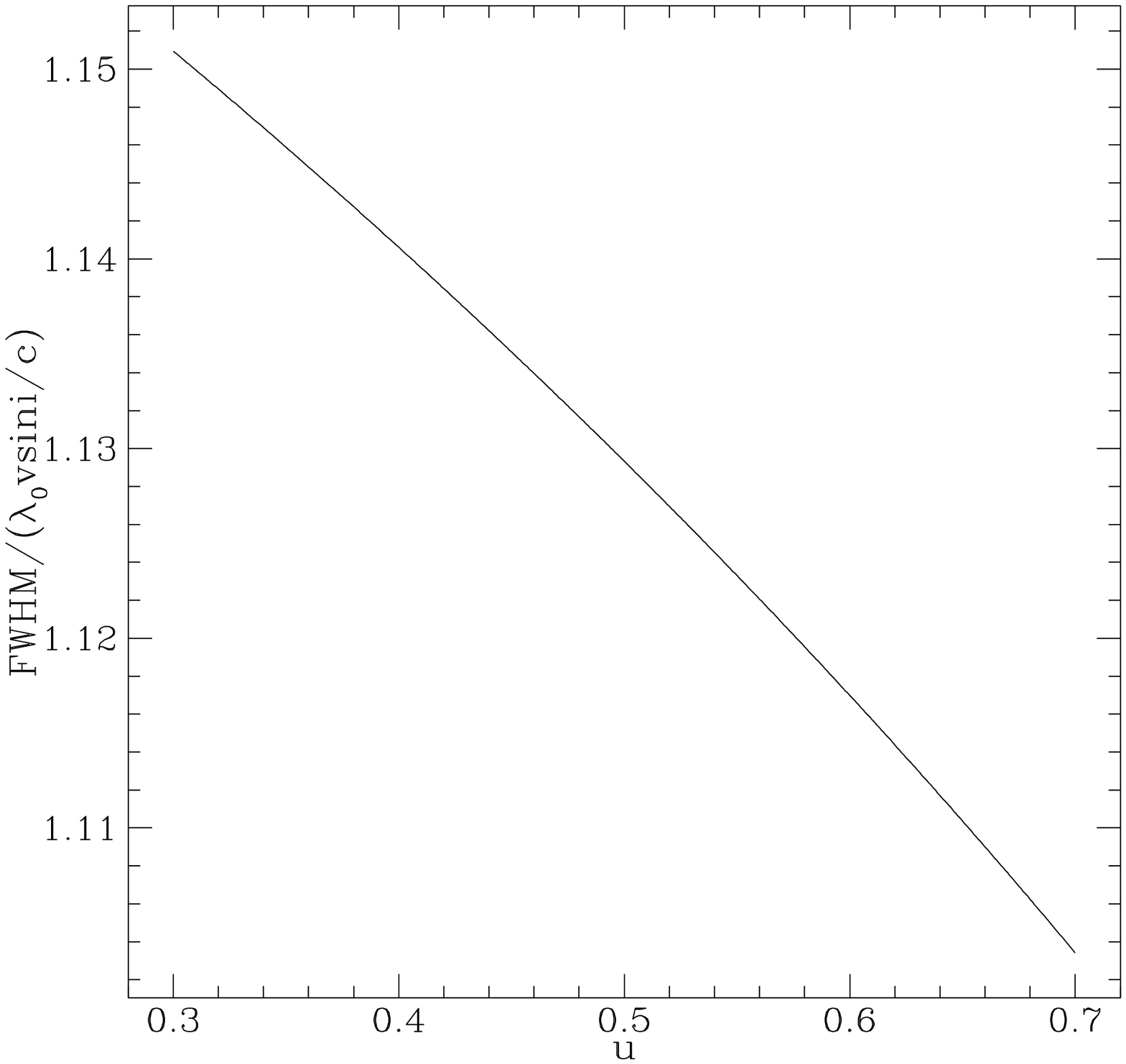,height=8cm}
\figure{1} {The value of the FWHM of an infinitely sharp
line which has been broadened by rotation as a function of the linear
limb-darkening parameter $u$. The full-width-half-maximum is
normalized by $\lambda_0{\vsini\over c}$ (see Appendix A).}
\endfig

The widths of the metal lines are measured by fitting a Gaussian plus
a straight-line (pseudo) continuum to the spectral line. As much
continuum as possible is used and where necessary double Gaussians are
fitted. The fitting process returns the FWHM of the best fitting
Gaussian, which is a measure of the second moment of the line
profile. The second moments for different limb-darkening laws are
derived in Appendix A; because of our CCD-pixel-spacing we use the
formulas for $\ln\lambda$-spacing. Note that Collins \& Truax (1995)
give the half-width at half-maximum instead of the FWHM. The formal
errors on the derived FWHM and amplitude of the best fitting Gaussian
vary between $\sim 1$\% and $\sim 40$\%. These errors are influenced
by uncertainties in the continuum level and a line-depth (at
line-centre) of at least 4--5\% is required if both the FWHM and
amplitude of the best fitting Gaussian are to have errors of less than
5\%.

\begfig 8cm
\vskip -8cm
\psfig{figure=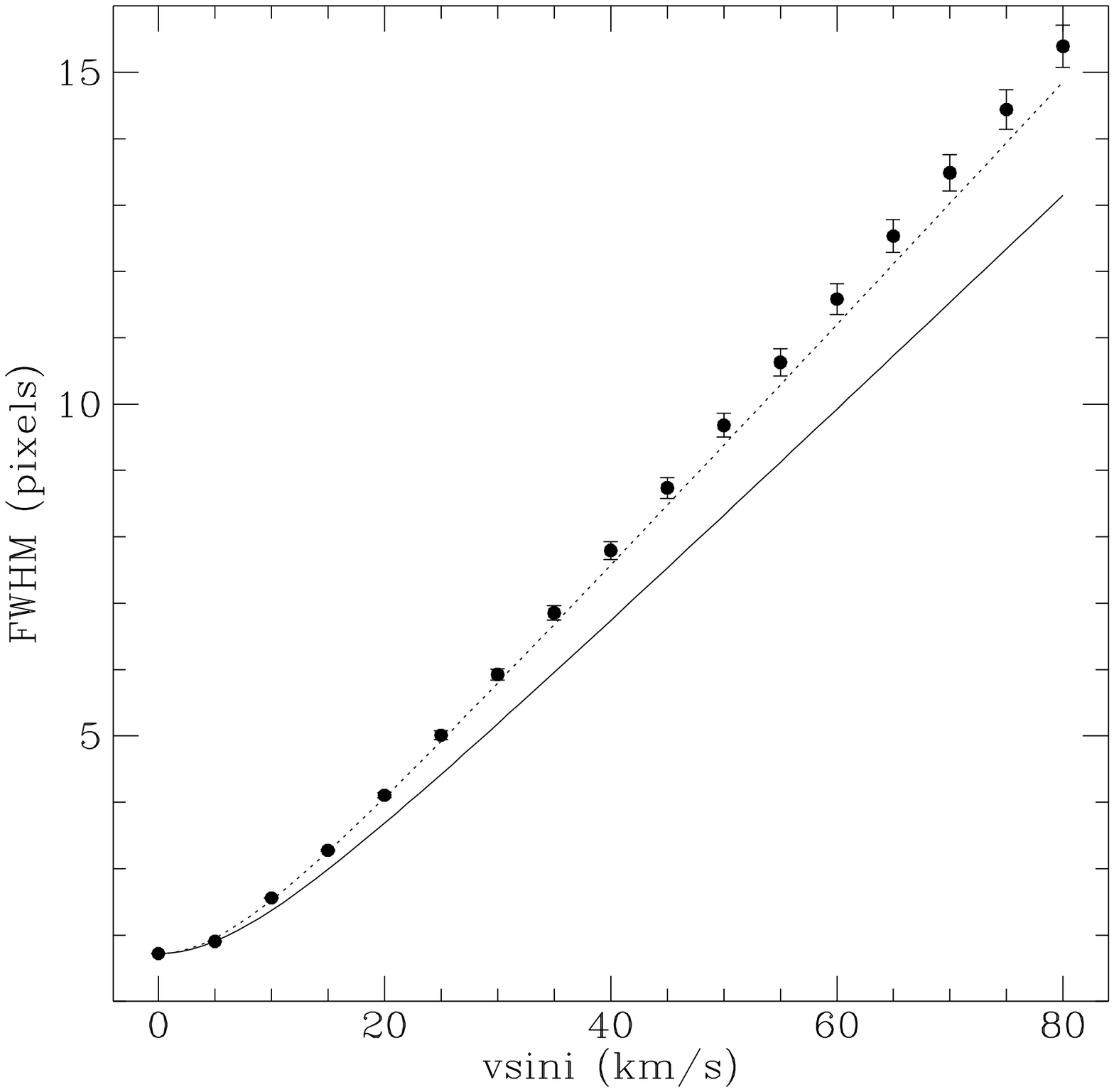,height=8cm}
\figure{2} {Analytically calculated FWHM of spectral
lines (solid curve) vs.\ the FHWM found by fitting a Gaussian
(dots). The results are shown as a function of \vsini\ for an
instrumental profile of 1.4 pixels FWHM, a thermal broadening of 1
pixel FWHM and $u=0.4$. The dotted line shows the result of fitting
FWHM$=\sqrt{a^2+b^2(\vsini)^2}$ to the dots. The fit is determined
primarily by the points at low \vsini. The error at high \vsini\
values amounts to 4\% at most.}
\endfig

The broadening of the lines depends on the assumed limb-darkening
law. We used a linear limb-darkening law to relate the line-widths to
\vsini. D\'\i az-Cordov\'es \ea\ (1995) give limb-darkening
parameters for the linear, quadratic and square-root approximations to
limb-darkening. Their values are based on Kurucz (1991) atmosphere
models. The value of the limb-darkening parameter for the linear law
varies from $\sim0.3$ to $\sim0.7$ for the range of spectral types in
our sample. We chose a constant value of $0.4$ (characteristic of
spectral type B3--B5, which is in the middle of our range) to derive
\vsini\ from the line-widths. Figure 1 shows the variation
of the FWHM as a function of the limb-darkening parameter. It is clear
from the figure that only small errors are made when using a single
value for $u$. The variation between the extreme values is $\sim 4\%$.

The measured line-widths also contain contributions from instrumental
and thermal broadening. These contributions have to be subtracted
before \vsini\ is derived. For our spectra, the instrumental
broadening amounts to 1.4 CCD-pixels FWHM for run E5 and 1.6 pixels
for the other runs. These values were derived from the measured widths
of thorium lines in the wavelength calibration spectra.  The only
metal lines that experience any significant thermal broadening (as
compared to the instrumental FWHM) are the C, N and O lines. The
amount of thermal broadening in these lines varies from $\sim
0.7$--$1.4$ pixels FWHM. We divided our spectral range roughly into
three temperature ranges. For stars with temperatures around
$\teff=8000~{\rm K}$ an average thermal profile of $0.7$ pixels FWHM
is used. For $\teff=15\,000~{\rm K}$ and $\teff=25\,000~{\rm K}$ the
values are $0.9$ and $1.2$ pixels FWHM. The differences are small, but
they are important for the stars with the sharpest lines. In practice
the spectral type ranges that correspond to the temperatures above
are: A1--B9, B8--B4, and B3--B0, respectively.

\begtabfull
\tabcap{1} {Values of parameters $a^2$ and $b^2$.}
\halign{\strut\hfil# & \quad\hfil#\hfil & \quad\hfil#\hfil
        & \quad\hfil#\hfil & \quad\hfil#\hfil \cr
\noalign{\smallskip\hrule\vskip2pt}
\omit&\multispan2\hfil E2/E3/E7\hfil&\multispan2\hfil E5\hfil \cr
\noalign{\vskip3pt}
\teff\hfil & $a^2$ & $b^2$ & $a^2$ & $b^2$ \cr
\noalign{\vskip2pt\hrule\vskip2pt}
8000 K & $3.05$ & $3.40\times10^{-2}$ & 2.45 & $3.35\times10^{-2}$ \cr
$15\,000$ K & $3.37$ & $3.40\times10^{-2}$ & 2.77 & $3.39\times10^{-2}$ \cr
$25\,000$ K & $4.00$ & $3.39\times10^{-2}$ & 3.40 & $3.40\times10^{-2}$ \cr
\noalign{\vskip2pt\hrule}}
\endtab

The fitting of Gaussians to estimate the line-widths introduces a bias
because of the non-Gaussian form of the rotational broadening
function. We have determined this bias by making artificially
broadened Gaussian lines that had intrinsic widths corresponding to
instrumental plus thermal broadening. The Gauss-fitting routine was
applied for a range of \vsini\ values and the results were represented
with a function of the form FWHM$=\sqrt{a^2+b^2(\vsini)^2}$. The
parameters $a^2$ and $b^2$ are then used in the derivation of
\vsini. Figure 2 shows an example of this investigation. The FWHM
values found with the Gauss-fitting routine are compared to the
analytically calculated values (Appendix A). The values for $a^2$ and
$b^2$ are listed in Table 1 for different stellar temperatures for
observing runs E2/E3/E7 and E5 separately.

\begfigside 9.6cm 12cm
\hskip -12.5cm
\psfig{figure=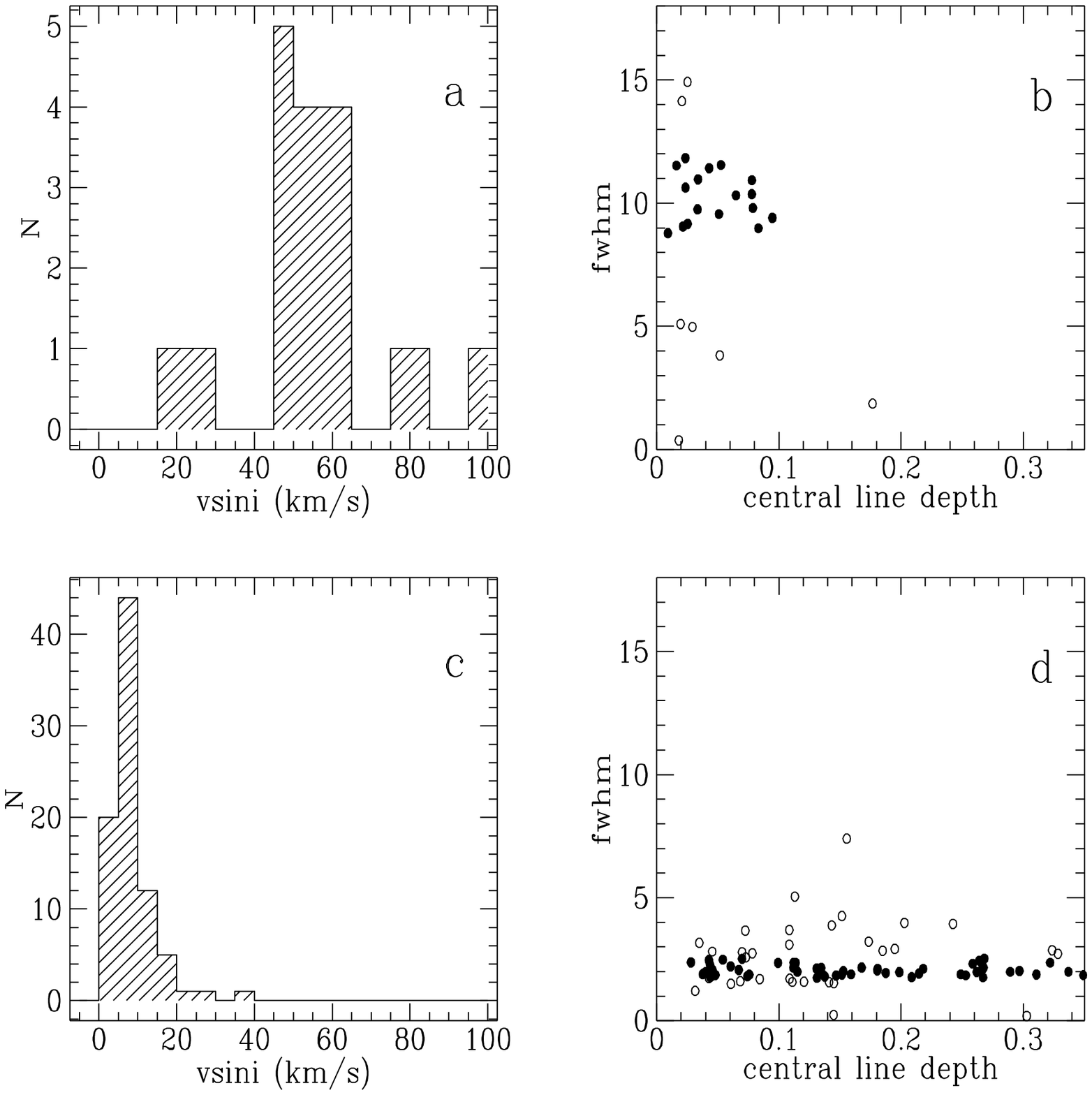,height=9.6cm,width=12cm}\hfil
\figure{3} {Histograms of \vsini\ values derived from
measured line-widths and the line-widths vs.\ line-strengths. The data
for HD 144218 (B2V) are shown in (a) and (b). The data for HD 72660
(A1V) are shown in (c) and (d). Indicated with solid circles are the
lines corresponding to the peaks in the histograms. For HD 144218 this
is the velocity range $45\kms\le\vsini\le 65\kms$ and for HD 72660 the
range is 0--$10\kms$ (the sharpest lines in HD 72660 formally give
negative \vsini\ values, after subtraction of thermal and instrumental
broadening, and are either interstellar, or affected by cosmic rays,
or under-sampled).}
\endfig

Finally, to convert the widths of metal lines to \vsini\ the following
procedure is used. The widths are measured for as many lines as
possible (typically 10--50, depending on \vsini\ and spectral
type) and converted to a histogram of \vsini\ values for each
star. This histogram will be contaminated by line-blends as well as by
interstellar lines. By selecting the bins where most of the
\vsini\ values are concentrated one can weed out the blended lines,
which cause
\vsini\ to be overestimated, and interstellar lines, which lead to
underestimates. This method has been used before by Day \& Warner
(1975). In practice it works best for sharp lined stars with a large
number of lines. One can then easily distinguish a peak in the
histogram. Examples are shown in Fig.~3 for two stars. One
is a sharp-lined A1V star ($\vsini=7\kms$) for which $\sim 130$ lines
were measured. The other is a B2V star rotating with
$\vsini\approx60\kms$, and for which 25 lines were measured. Also
shown is the FWHM vs.\ line-strength for all lines measured. The lines
that correspond to the peaks in the histograms are indicated as solid
circles. Note that these lines define a definite locus and that for
these two examples there is no increase of the FWHM with
line-depth. All other lines are either blends, interstellar, or
affected by cosmic rays. In the case of HD 144218 the value of \vsini\
is higher and the peak in the FWHM distribution is less well defined.

Once the most likely range for \vsini\ has been selected from the
histograms we proceed by checking whether there is a significant
correlation between line-depth and FWHM. Such a correlation is
expected if lines are on the flat part of the curve of growth, where
the FWHM increases with line-depth. If a significant correlation is
found we correct the FWHM of each line by fitting a linear model to
the data. After this correction we check whether the distribution of
the FWHM of the lines is significantly skewed towards high values. If
this is the case there may still be line blends in the data. We
correct for these blends by repeatedly discarding the line with the
largest FWHM until the rms spread in the data changes by no more than
10\%. This procedure is applied subject to the condition that no more
than 30\% of the lines are discarded. The two quantitative criteria
were found by trial and error. After the selection procedure is
completed \vsini\ is calculated as a weighted average of the
\vsini\ values given by each line. In order to ensure that lines with
erroneously low formal errors on their FWHM do not enter into this
average with too much weight, the median error in the FWHM is
calculated from the data and half of this value is added in quadrature
to all errors.

\titleb{Deriving \vsini\ by Template Broadening}
For stars with \vsini\ larger than about $50\kms$ the method for
deriving \vsini\ described above becomes unreliable. This is because
the estimated stellar continuum level is already significantly too low
due to rotational broadening, which implies that the measured FWHM of
the lines is underestimated. Furthermore, at higher values of \vsini\
the number of lines for which one can measure the FHWM is reduced, and
it is more difficult to identify and eliminate line blends. Therefore,
a different technique for obtaining \vsini\ is required.

The technique we consider here is that of broadening of spectra of
sharp-lined stars and comparing the broadened spectrum to the spectrum
of the target star for which \vsini\ is to be determined. Again, the
model that is used when broadening a template spectrum is the
classical model for rotating stars. In this case all three assumptions
of the classical model play a role. The assumptions on limb-darkening
are now very relevant because of the inclusion of broad and strong
hydrogen and helium lines in the determination of \vsini. The
limb-darkening we used was again linear limb-darkening with a
parameter $u=0.4$. The effects of varying the limb-darkening are
discussed at the end of Appendix C. At the highest \vsini\ values the
deformation of the target star, which leads to changes in line-shapes
over its surface, shows up as a mismatch between its spectrum and the
broadened template spectrum (see also Appendix C). The template stars
were taken from our own sample and they were required to have a low
value of \vsini\ ($\sim10$--20\kms), which was determined using method
1.

Bearing all the mentioned caveats in mind we proceed as follows. We
broaden a template spectrum for a number of values of \vsini\ over a
chosen interval and find the value for which the difference between
broadened template and target is minimal. The search interval is
chosen based on the literature value for \vsini\ or by visual
inspection of the spectrum of the target. The comparison between the
broadened template spectrum and the target spectrum is done for 10
orders in the ECHELEC spectra (i.e., orders 139--148, see Verschueren
\ea 1996). To find the best \vsini\ value a figure of merit analogous
to a $\chi^2$ is constructed for each order separately. This figure of
merit is the mean absolute difference, over all pixels, between the
template and the target divided by the noise in the target
spectrum. This is essentially an automation of the comparison that one
would do by eye. Based on the values of the figures of merit a
weighted average of the \vsini\ values found for each order is
calculated. Values of
\vsini\ at the edge of the search interval are discarded. These values
usually point to noisy orders, orders with low information content or
orders where there is a significant mismatch between the spectra of
the template and the target. In all cases the values of \vsini\ found
for all orders are inspected to check whether the results are
consistent. The wrong choice for the search interval can cause a large
spread in the data that is not real.

The advantage of doing the comparison per spectral order is that it
provides information on \vsini\ independently for different parts of
the spectrum. This provides an effective way of detecting mismatches,
either intrinsic or due to the data reduction process, between the
template and target spectra by demanding consistency of \vsini\
values. A global comparison only gives a single value of
\vsini\ and spectral type mismatches are then only revealed by
visual inspection. Furthermore, from the spread in the data and a
visual inspection of the results one gets a good indication of the
precision as well as the accuracy with which \vsini\ was
determined. Finally, there are often weaker lines present in the broad
wings of the Balmer lines. The influence of these lines is not taken
into account in the models by CTC. These lines can make a difference
if they are strong enough and template broadening automatically takes
their contribution into account.

Spectral type mismatches often result from slight differences in
temperature or luminosity class. The temperature differences are
readily apparent in the strength of the helium lines for B-stars and
the luminosity differences are manifest in the hydrogen lines. Another
problem to take into account when looking for a template spectrum is a
possible error in spectral classification of either template or
target. The only way to get around this problem is to try different
template spectra and find the best match.

\begfigside 14.85cm 12cm
\hskip -12.5cm
\psfig{figure=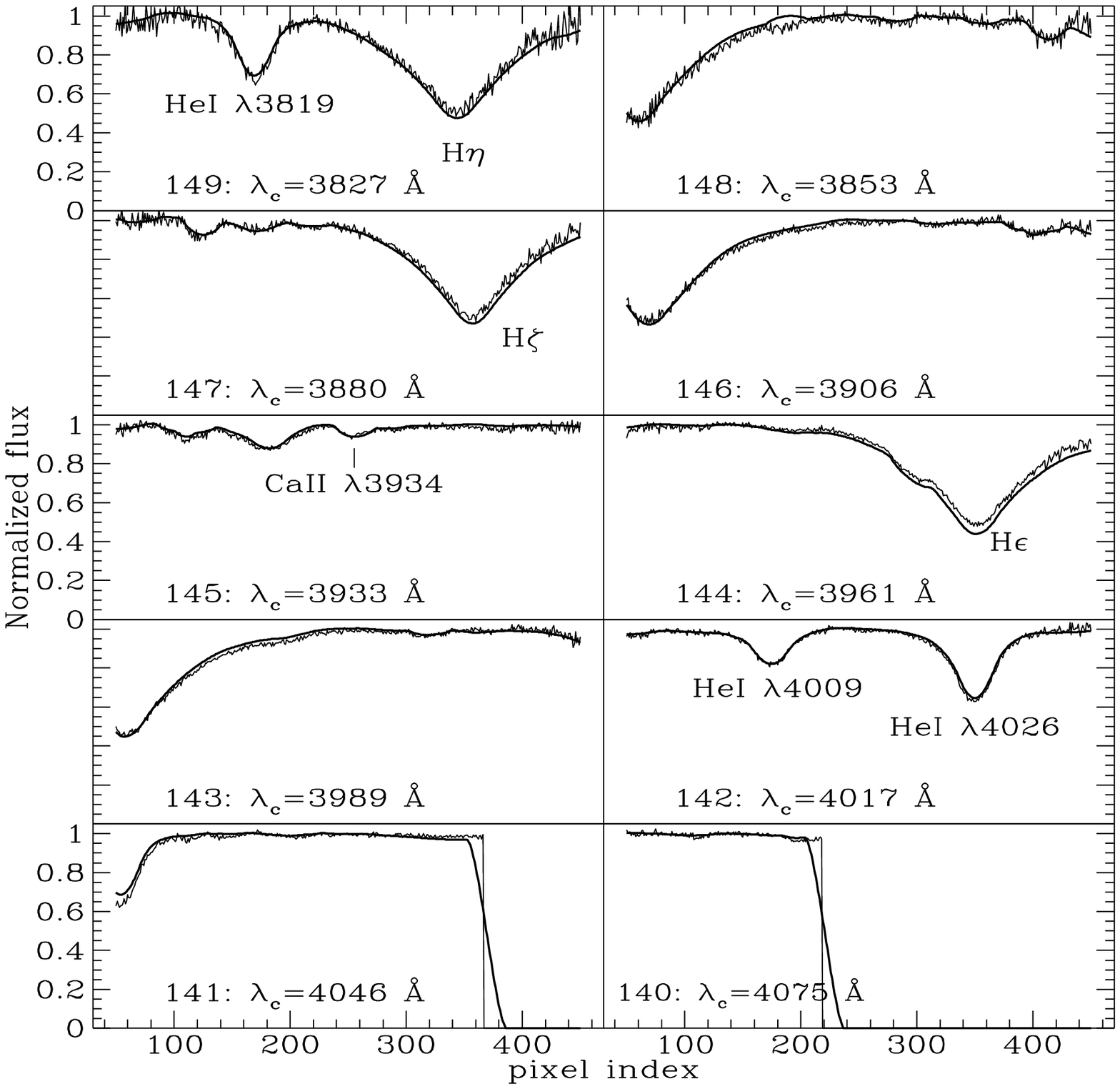,height=14.85cm,width=12cm}\hfil
\figure{4} {Comparison of the target spectrum of HD 121790
(B2IV-V) with the best-fit broadened template spectrum of HD 96706
(B2V). The thick line is the broadened template spectrum. The central
wavelength of each order is indicated, as well as several spectral
lines. The best value of \vsini\ for the target star is
$118\kms$. Note the mismatches between the spectra of the stars at
several places in the continuum as well as in the wings of the
hydrogen lines. These mismatches can be due to intrinsic spectral type
mismatches or due to the data reduction process.}
\endfig

\begfigside 14.85cm 12cm
\hskip -12.5cm
\psfig{figure=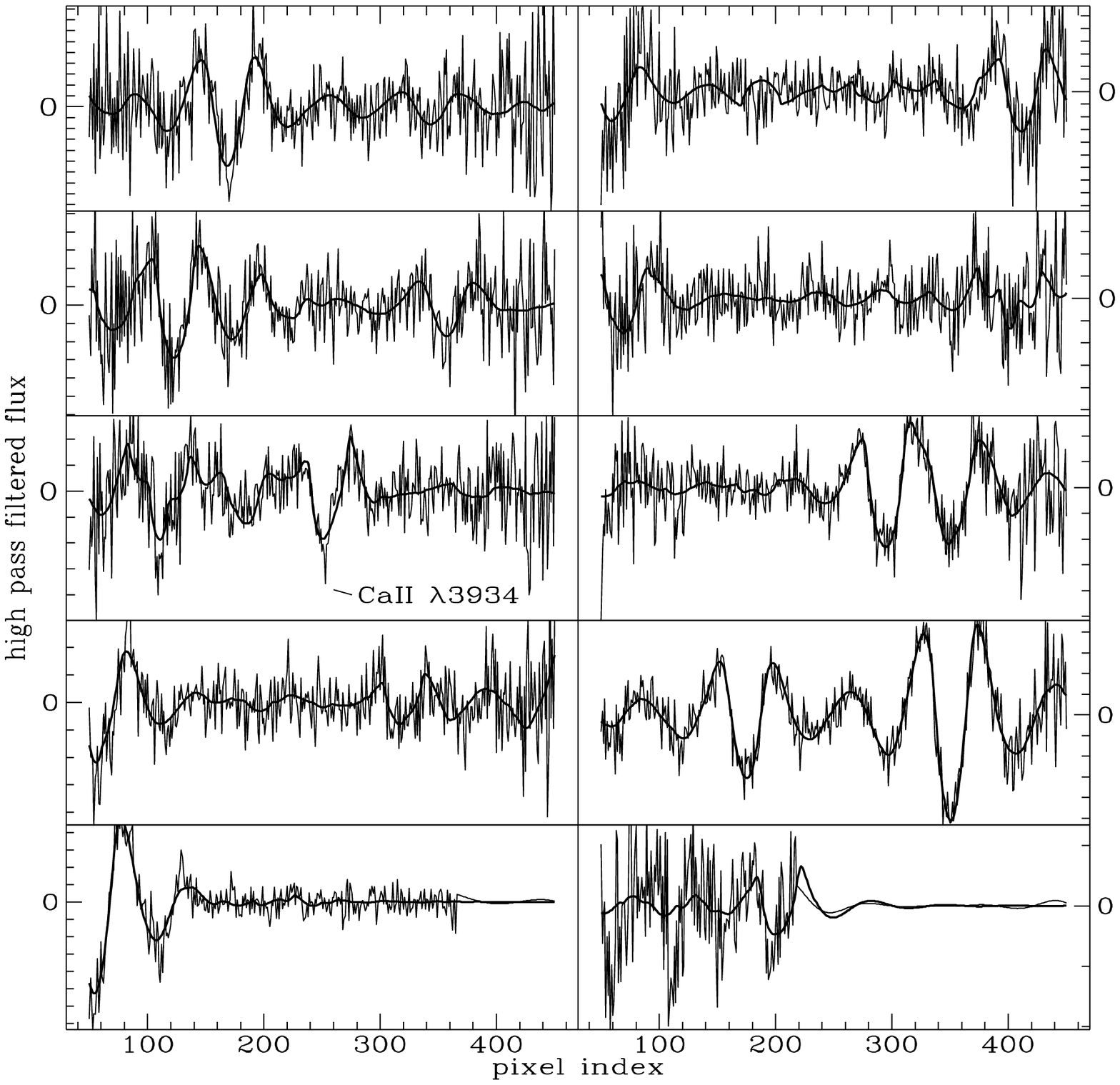,height=14.85cm,width=12cm}\hfil
\figure{5} {The same as Fig.~4 but now for the
high-pass filtered spectra. The order numbers and central wavelengths
are the same as in Fig.~4. The distance between the tick-marks
on the vertical scale is $0.01$. The high-pass filter takes out all
spatial wavelengths longer than $358\kms$ (see also appendix
B). Notice the very good fit of the two spectra.}
\endfig

The information on \vsini\ in the rotationally broadened spectra
resides in the metal lines and in central parts of hydrogen and helium
lines. It is important to get rid of the continuum and of the broad
wings of Stark-broadened lines when calculating the figure of merit,
otherwise these components dominate the determination of \vsini,
whereas in practice they carry no information on \vsini. This problem
can be solved by high pass filtering both spectra before comparing
them. The filtering is done with a high-pass frequency filter in
Fourier-space. The details of the filtering process are given in
Appendix B.

Figures 4 and 5 show an example of the fitting
process for a star for which the best value of \vsini\ is 118\kms. It
is HD 121790, a B2IV-V, star and the template is HD 96706 with
spectral type B2V. Figure 4 shows a direct comparison of the
broadened template and the target spectrum. Figure 5 shows a
comparison of the Fourier filtered spectra. All spatial wavelengths
longer than $358\kms$ were filtered out in this case. Note that in
Fig.~4 there are several mismatches between the two spectra
in the continuum as well as in the wings of the hydrogen lines. These
mismatches bias the determination of \vsini\ if no filtering is
applied. The comparison between the two filtered spectra shows
convincingly that the broadened template spectrum fits the target
spectrum. The interstellar \ionII{Ca}3934\ line is located in the
central part of order 145. This line is left out of the comparison
between the spectra.

As mentioned in the beginning of this section, the template broadening
technique fails at high values of \vsini. The reason is that the
classical model for rotational broadening does not take the effects of
gravity darkening into account. For stars rotating at a large fraction
of their break-up velocity the surface gravity decreases in the
equatorial regions. This leads to obscuration of the parts of the
observed line profile that come from the equatorial regions. The
result is that rapidly rotating stars which are observed equator-on
have line profiles that resemble those of more slowly rotating
stars. Furthermore, because of the strong lines that are used we
expect the effects of limb-darkening to be more severe. We
investigated both these effects on the template broadening method by
using synthetic data generated from the models by CTC. These
investigations are described in Appendix C. The conclusions are that
the choice of the limb-darkening law is not important for template
broadening, but gravity-darkening is important and one should be
careful when using template broadening results as soon as \vsini\ is
larger than $\sim 120$--$150\kms$.

\titleb{Determining \vsini\ for Rapid Rotators by Model Fitting}
For the stars with the highest values of \vsini\ one is only left with
the alternative of comparing model line profiles to the data. The only
line in our spectrum for which rotating stellar models have been
calculated is the \ionI{He}4026\ line. The models are given by CTC for
a set of values of $w$, the fraction of the critical angular velocity
at which the star rotates, and $i$, the inclination angle of the
rotation axis with respect to the line of sight. The critical angular
velocity of the star corresponds to the equatorial rotational velocity
at break-up. The models were calculated for main sequence stars and
for stars of luminosity class III for the spectral types B1, B3, B5,
B7, and B9. The CTC models are based on the Roche model described in
Collins (1963, see also Sect.~4.2). The basic procedure we follow is
to compare the whole line profile to the data for each $(w,i)$
combination and determine \vsini\ from the best fitting models. We
stress here that we cannot derive $w$ and $i$ separately with our
data. Usually a number of $(w,i)$ combinations are consistent with the
data and one can only derive \vsini\ with a corresponding range of
uncertainty.

From the set of models given by CTC we constructed a grid of models
for each spectral type. In this grid $w$ varies from 0 to $1.0$ in
steps of $0.1$ and $i$ varies from \deg{0} to \deg{90} in steps of
\deg{10}. The models at grid points not given by CTC were constructed
by linear interpolation between CTC models. We verified that using
higher order interpolation methods does not improve the model grid. We
also constructed models for B4, B6 and B8 stars by linear
interpolation between spectral types given by CTC. A similar procedure
is not possible for B2 stars because they have the strongest
\ionI{He}4026\ lines; and here interpolation between B1 and B3 models
is incorrect. For B2 stars B3 models were used and we discuss the
implications below.

\begfigside 10.5cm 12cm
\hskip -12.5cm
\psfig{figure=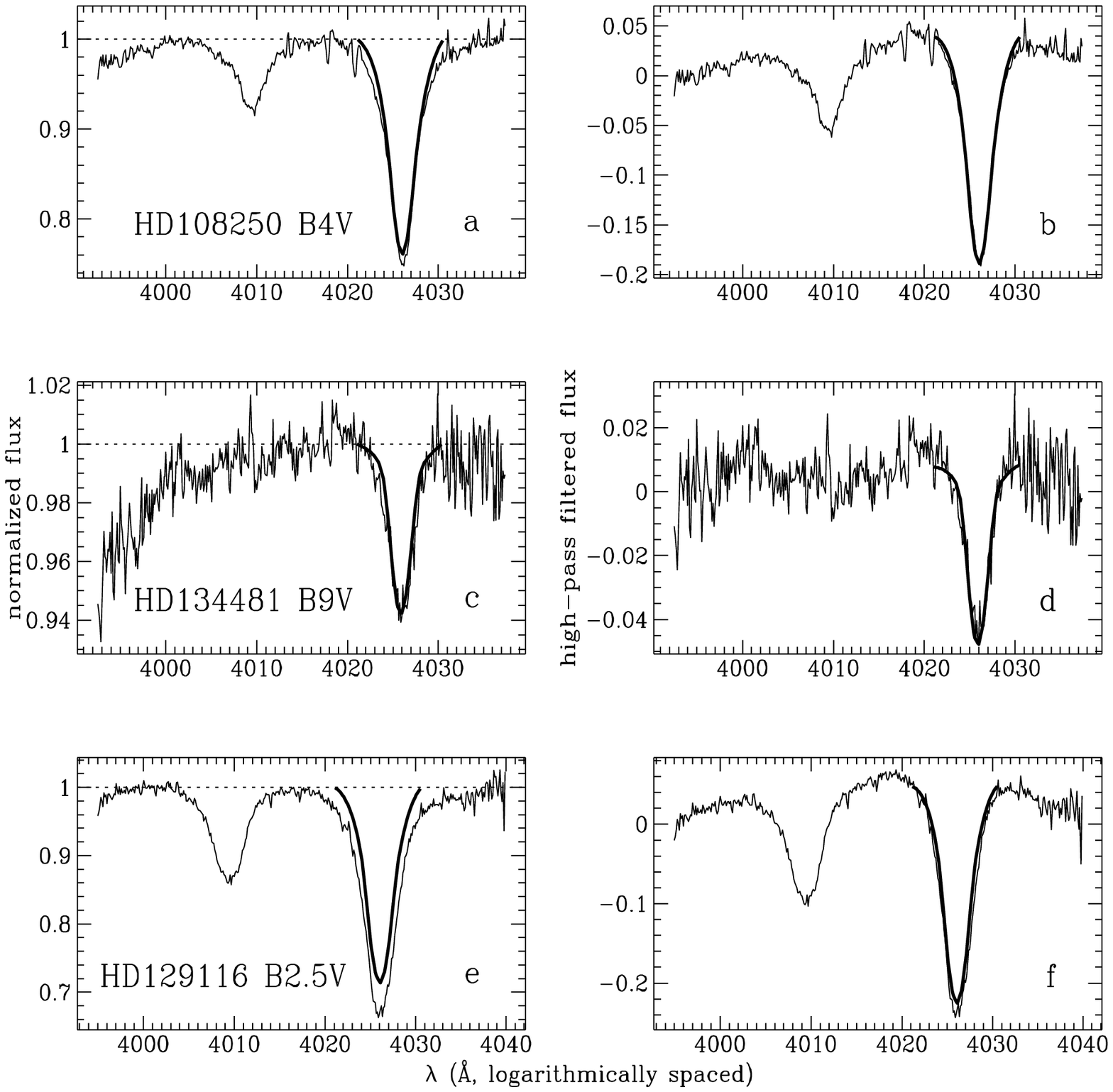,height=10.5cm,width=12cm}\hfil
\figure{6} {Three examples of the model fitting
process for deriving \vsini. Figure (a) shows a direct comparison of
the best fitting model (thick line; $w=0.6$, $i=\deg{50}$,
$\vsini=130\kms$) and the data for HD 108250. Figure (b) shows the
same as (a) after high-pass filtering. Figures (c) and (d) show the
same as (a) and (b) for HD 134481. The best fitting model in this case
is: $w=1.0$, $i=\deg{30}$, $\vsini=174\kms$. Figures (e) and (f) show
the same as (a) and (b) and illustrate the effect of using a B3 model
to fit the data of a B2 star. The \ionI{He}4026\ line is weaker for B3
stars than for B2 stars and the effect is that
\vsini\ is underestimated. The best fitting model in this case
is: $w=0.5$, $i=\deg{60}$, $\vsini=124\kms$. The dashed lines indicate
the continuum level of the models and illustrate the need for
high-pass filtering.}
\endfig

The model profiles are given as a function of wavelength so we used
spectra rebinned into $\ln\lambda$-space. The \ionI{He}4026\ profiles
were transformed to $\ln\lambda$-space (spacing in $\ln\lambda$:
$2.8\times10^{-5}$).  The data and the models were compared for each
$(w,i)$ pair and the pair for which the reduced $\chi^2$ is at a
minimum was considered to correspond to the best fitting
model. Possible radial velocity shifts were taken into account by
automatically shifting the model with respect to the data for each
$(w,i)$ pair and searching for the minimum $\chi^2$.

In order to estimate the uncertainties in the derived value of \vsini\
the ten models with the lowest $\chi^2$ were considered. Of these, a
model was considered equivalent to the best fitting model if the
associated $\chi^2$ did not differ by more than a factor of 2 from the
minimum and if the emergent flux at all points in the model did not
differ by more than $0.01$ for \ionI{He}4026\ lines with a central
depth of $0.9$ or less. The maximum tolerated difference in fluxes was
$0.005$ for \ionI{He}4026\ lines with a central depth of more than
$0.9$. The \vsini\ values corresponding to the best fitting models
were subsequently averaged with the corresponding $\chi^2$ values as
weights. The rms spread in \vsini\ values was taken as the error
bar. In some cases one finds only one best fitting model and then the
adjacent points located between the grid points were considered in
order to assign an error bar to \vsini.

As was the case with template broadening, there may be mismatches
between model and data due to the continuum. We turned to high pass
filtering again to get around the continuum terms. Because one wants
to compare the whole line profile to the data, and because most of the
target stars have a high value of \vsini, only the most important
continuum terms should be filtered out. In order to safely examine
stars with values for \vsini\ up to 500\kms, we chose to filter away
all spatial wavelengths larger than $\sim1000\kms$. None of the stars
in our sample have \vsini\ values this high.

Figure 6 shows three examples of the model fitting process. In all
cases the comparison between data and best fitting model is shown both
for the high-pass filtered data and for the un-filtered data. In
Figs.~6a and 6b the data for HD 108250 (B4V) are shown. The best
fitting model is indicated by the thick line. Note that in the direct
comparison it appears as if the \ionI{He}4026\ model line is not deep
enough and that the wings of the line profile do not match the data
either. This is due to the continuum in the data which has not yet
been properly rectified. This can be seen from Fig.~6b; the Fourier
filtered model and data now match well in both the line core and in
the wings. Figures 6c and 6d show the same for HD 134481 (B9V). In
this case the continuum close to the
\ionI{He}4026\ line just happens to have the right value.

The strength of the \ionI{He}4026\ line reaches a maximum around
spectral type B2--B3 and the B3 models are not always strong enough
compared to the data. This is often the case when one considers B2
stars. Figures 6e and 6f show the effects of using B3 models to fit
the data for a B2 star (here HD 129116 B2.5V). The
\ionI{He}4026\ line is weaker in the B3 model than in the B2 star. Because
both the broader wings and the deeper line core of the B2 star have to
be fitted, \vsini\ is underestimated.

All results were inspected by eye to judge whether the best fitting
model was actually correct. We encountered stars for which the models
corresponding to their spectral types did not fit the data. In those
cases other models were tried and the spectral type for which the best
fit was obtained was used to derive \vsini. For the majority of these
cases the strength of \ionI{He}4026\ and the ratio of \ionI{He}4026\ to
\ionI{He}4009\ in our spectra indeed were not
consistent with the spectral type of the star.

Finally, we remark that the \vsini\ values are based on the rotational
velocity of the star at the theoretically calculated break-up
velocity, and the value of $w$ and $i$. This means that the later
interpretation of \vsini\ distributions in terms of break-up
velocities will be complicated (see Sect.~4).

\titleb{Results}
The results of the \vsini\ determinations are shown in Tables
2a--2d. The tables contain the following data. Column (1) gives the HD
number of the star and column (2) gives the number of the star in the
HIPPARCOS Input Catalogue (Turon \ea 1992). In column (3) the
spectral type is given as listed in the HIPPARCOS Input Catalogue. The
derived value of the projected rotational velocity and its associated
error are listed in columns (4) and (5). For stars for which we have
\vsini\ values from different runs the error-weighted average is
listed. For the following sharp-lined stars the \vsini\ values from
two or more runs are not consistent within the error bars: HD 109668,
HD 110879, HD 120307, HD 122980, HD 132955, HD 133955, HD 145502,
HD 147165, HD 158408, HD 172910. These stars all turn out to be
spectroscopic binaries or radial velocity variables (see below). CTC
provide a value for the break-up velocity (\vbreak) for each spectral
type for which model profiles were calculated.  Columns (6) and (7)
list this break-up velocity and the ratio $\vsini/\vbreak$. In cases
where the CTC model profiles were used the value of \vbreak\
corresponds to the spectral type of the best fitting model and not
necessarily to the listed spectral type (see also the notes to Table
2). For stars later than A0 and for supergiants we determined the
break-up velocities based on stellar parameters given by Strai\v zys
\& Kuriliene (1981) and Harmanec (1988). For HD 149757 and HD 151804
we took the mass and radius given by Lamers \& Leitherer (1993). In
column (8) we list which method was used to derive \vsini. The method
is indicated by its number and for some stars more than one method was
used and a weighted average of \vsini\ was then formed. For a number
of stars none of the methods for determining \vsini\ works (these are
very early-type stars or late-type stars with only blended lines, or
spectroscopic binaries for which line-profile fitting does not
work). For these stars \vsini\ was estimated by visual inspection of
the spectrum or by taking a literature value; as indicated in the
notes. In column (8) this is indicated by the number 4. Column (9)
indicates whether or not a star is a member of Sco OB2; 'm' indicates
that the star is a photometric member, as found by de Geus \ea (1989),
and 'M' indicates that the star is a proper motion member, as found by
Blaauw (1946) and Bertiau (1958). Finally, in column (10) we list
whether the star has a constant radial velocity (CON), is a radial
velocity variable (RV), a single-lined spectroscopic binary (SB1) or a
double-lined spectroscopic binary (SB2). For stars without an entry in
column (10) there is no good information on their binarity.

The duplicity information was obtained from the Bright Star Catalogue
(Hoffleit \& Jaschek 1982; Hoffleit, Saladyga \& Wlasuk 1983), Levato
\ea (1987) and from preliminary findings by Verschueren
\ea (1996) based on our data. A star is considered a radial velocity
variable if it is listed as SB or SB1 in the Bright Star Catalogue or
if Levato \ea or Verschueren \ea list it as a radial velocity
variable. A star is considered a single-lined spectroscopic binary if
the Bright Star Catalogue lists it as SBO or SB1O or if Levato
\ea determined an orbit and list it as SB1. A star is considered a
double-lined spectroscopic binary if any of these references list it
as SB2 or SB2O. Note that the duplicity information pertains to
duplicity that is apparent in the spectrum of the observed object;
either through variability of the radial velocity or because of the
presence of double lines. Thus, although some stars are part of a
(wide) binary or multiple system, if there are no signs of duplicity
in their spectrum they are listed as single stars. Finally, an
asterisk after column (10) indicates that there is a note about that
star in Appendix D. The notes contain information on the derivation of
\vsini. They also provide information about non-radial pulsations in
the stars. These pulsations can have a sizeable effect on the line
profiles and thus on the derived value of \vsini. The information on
the non-radially pulsating stars in our sample (including the $\beta$
Cephei variables) was kindly provided by P.\ Groot.

\begtabfull
\tabcap{3} {Projected rotational velocities for
early-type stars not related to Sco OB2.}
\halign{\strut\hfil# & \quad#\hfil & \quad\hfil#
            & \quad\hfil#\hfil & \quad\hfil#\hfil \cr
\noalign{\smallskip\hrule\vskip2pt}
HD\hfil & \hfil MK & \vsini\hfil & $\sigma(\vsini)$ & Dupl.\cr
\noalign{\vskip2pt}
\omit & \omit & $\kms$ & $\kms$ & \omit \cr
\noalign{\vskip2pt\hrule\vskip2pt}
28114  & B6IV        &  27 &  4 & \omit \cr
35039  & B2IV-V      &  11 &  3 & \omit \cr
36591  & B1V         &  10 &  3 & \omit \cr
45572  & B9V         &  20 &  3 & \omit \cr
57682  & O9IV        &  24 &  5 & \omit \cr
72660  & A1V         &   7 &  2 & \omit \cr
91316  & B1Iab       &  66 &  6 &  SB1  \cr
179761 & B8II-III    &  16 &  3 & \omit \cr
\noalign{\vskip2pt\hrule}}
\endtab

For the star HD 140008 it was possible to measure \vsini\ for both
components in the spectrum, and to within the error bars they both
have the same \vsini\ of $11\kms$. Both components were taken into
account in the subsequent analysis of the data. For method 2 the
following stars were used as templates: HD 96706, HD 120709,
HD 126341, HD 129056, HD 132955, HD 134687, HD 149438. In Table 3 we
give the projected rotational velocities of the additional set of
early-type stars that was observed with ECHELEC. Of these stars the
following have been used as templates for method 2: HD 28114,
HD 35039, HD 36591, HD 45572, HD 72660, HD 179761.

Finally, the stars HD 140008, HD 143118 and HD 144294 are listed as
members of UCL by de Geus \ea (1989), but in order to be consistent
with the subgroup boundaries as given by de Zeeuw \ea (1994) we list
these stars as members of US.

\titleb{Consistency of the Three Methods for Determining \vsini}
Figure 7 shows a comparison of the \vsini\ values
derived using the three methods described above. In
Fig.~7a the results from line-width measurements and
template broadening are compared. All the values are consistent within
the error bars. However, the values from line-width measurements are
systematically lower, although generally this effect is $\lessapprox
3\kms$. This is an effect of the lowering of the apparent continuum
due to rotational broadening. A lower continuum means that the
measured line-widths are too small, which leads to underestimates of
\vsini.

\begfig 8.8cm
\vskip -8.8cm
\psfig{figure=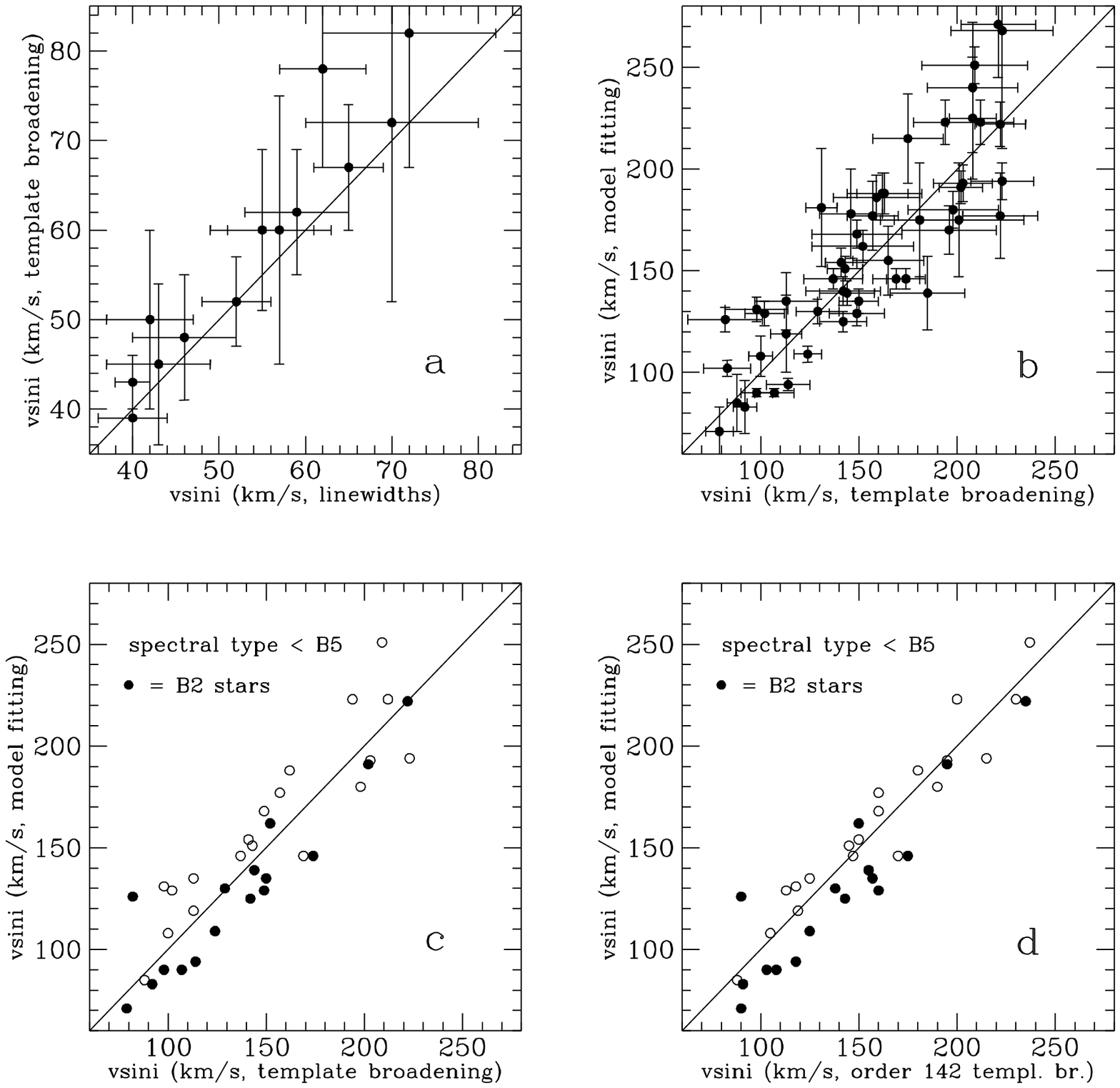,height=8.8cm}
\figure{7} {Comparison of \vsini\ values obtained
from measurements of line-widths, template broadening and fitting
models to the \ionI{He}4026\ line. Figure (a) shows the comparison
between the line-width method and template broadening. Figure (b)
shows a comparison of the results from template broadening and model
fitting and (c) shows the same for stars of spectral type earlier than
B5. The B2 stars are indicated as black dots. Figure (d) shows the
same as (c) but now the template broadening values are derived from
order 142 (containing the \ionI{He}4026\ line) only. See the text for
discussion.}
\endfig

Figure 7b shows the comparison of \vsini\ values derived
from template broadening and fitting the \ionI{He}4026\ line with
models. Template broadening was applied for all spectral types over a
large range of \vsini. Thus, bearing in mind the discussion in
Sect.~3.2 and Appendix C, we expect the two methods to yield
inconsistent results for $\vsini\largapprox150\kms$.  This can be
clearly seen in Fig.~7b. For 17 out of the 52 stars in
this figure the \vsini\ values from each method are inconsistent if
one takes the error bars into account.

In order to understand Fig.~7b better we break the plot
down according to spectral type. For spectral types earlier than B5
most stars have $\vsini/\vbreak\lessapprox0.5$. The opposite is the
case for later spectral types. So for later spectral types $w>0.7$
(see Sect.~4 for a discussion on the relation between
$\vsini/\vbreak$ and $w$). That is precisely the range where template
broadening fails. Figure
7c shows the comparison of the two methods if one leaves
out spectral type B5 or later. The scatter at the high \vsini\ end is
much reduced but overall it is still significant. We have also
indicated the B2 stars separately and one can see that most of them
have model fitting values that are lower. We noted in the previous
section that this is to be expected if the B3 \ionI{He}4026\ model lines are
not strong enough compared to the data.

Figure 7d shows the comparison if one uses only the results for
template broadening from order 142 (containing the
\ionI{He}4026\ line). Most of the B2 stars are displaced downwards, which
is consistent with the remarks about B3 models made above. The other
stars now lie closer to the line of consistent values and overall the
scatter is somewhat reduced. The results in this figure suggest that
the Balmer lines in other ECHELEC orders lead to underestimates of
\vsini\ for template broadening. This is indeed confirmed if one
compares the template broadening results of all other ECHELEC orders
to those from order 142.

Another cause for inconsistencies between template broadening and
model fitting comes from the fact that one has to make assumptions
about the stellar parameters in model fitting. One needs to know the
radius and mass of the stars in order to calculate \vbreak\ and hence
\vsini. This problem does not arise when one uses the classical model
for rotational broadening. For example: CTC give a mass of $11\msun$
and a radius of $5.8$~R$_\odot$ for a B1V star. However, according to
the calibrations by Strai\v zys \& Kuriliene (1981) a B1V star has a
mass of $12.9\msun$ and a radius of $5.9$~R$_\odot$, while Schaller
\ea (1992) use $9\msun$ and $3.7$~R$_\odot$. These parameters lead to
break-up-velocities of 491, 527 and 556\kms, respectively. So a
variation of more than 10\% in \vsini\ is possible due to
uncertainties in stellar parameters.

We conclude that the three methods for obtaining \vsini\ are on the
same velocity scale. However, caution should be exercised when
comparing the results from template broadening to those of model
fitting. The average difference between line-width and template
broadening results is 7\%, and for template broadening and model
fitting this difference is 10\%.

\begfig 8cm
\vskip -8cm
\psfig{figure=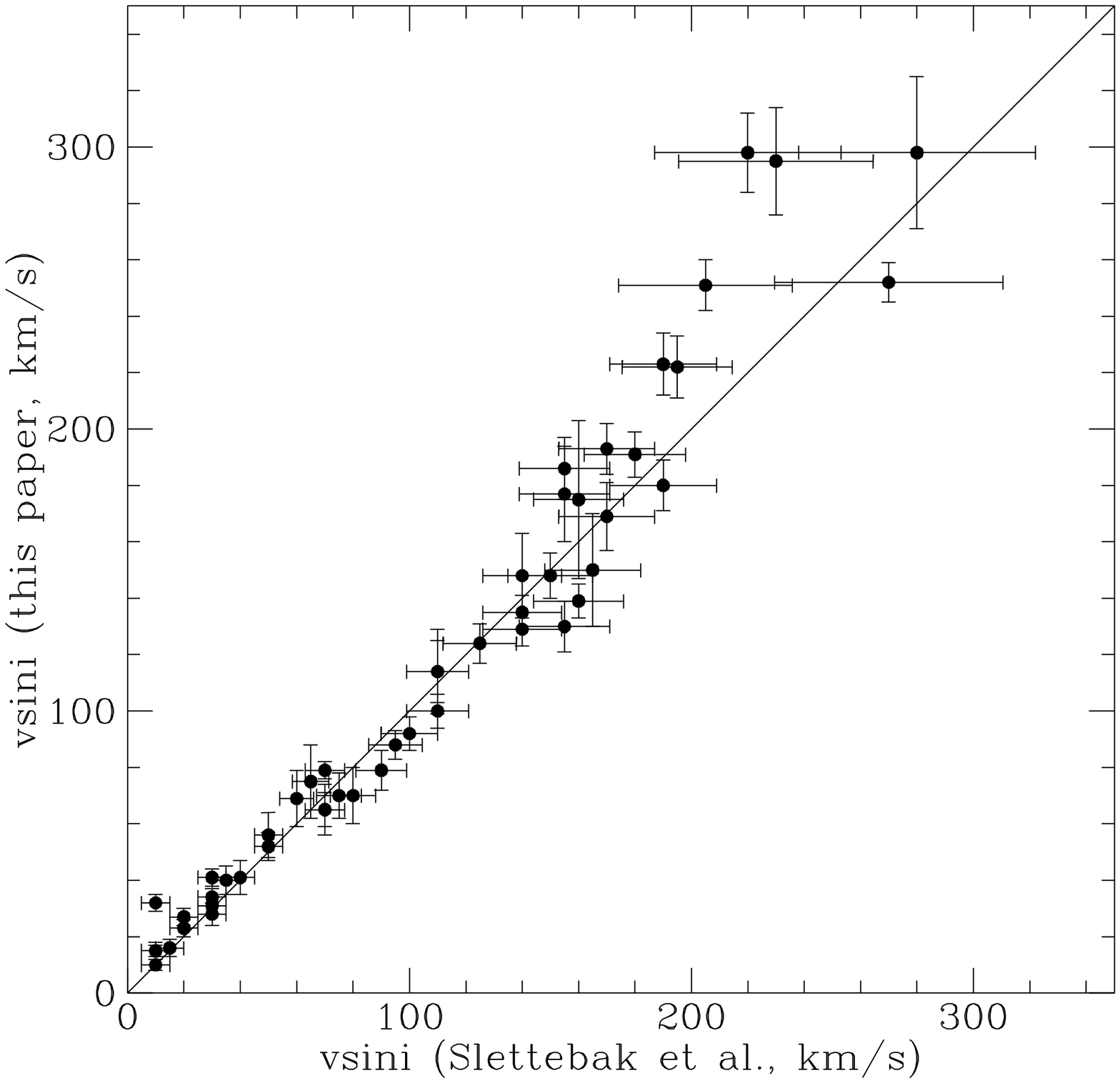,height=8cm}
\figure{8} {Comparison of the \vsini\ values derived by
SCBWP and the values derived in this paper.}
\endfig

\titleb{A Comparison of \vsini\ Values for Standard Stars}
We now turn to the comparison between our results and the values for
\vsini\ listed for the standard stars in the system presented by
Slettebak \ea (1975, hereafter SCBWP). They determined projected
rotational velocities for stars with a range of spectral types and
\vsini. These measurements define a \vsini\ velocity scale to
which other measurements can be compared.

Figure 8 shows the comparison between \vsini\ derived by
us and the values listed in SCBWP (1975). We have 51 stars in our
sample in common with SCBWP. The error bars on the values of SCBWP are
assigned according to the accuracy given in their paper (10\% for
$\vsini\lessapprox200\kms$ and 15\% for $\vsini\largapprox200\kms$),
where for values of \vsini\ less than $50\kms$ an error bar of $5\kms$
is assigned. Within the error bars almost all \vsini\ values are
consistent, but below about $60\kms$ most of our values are
systematically offset from those by SCBWP to higher \vsini. The values
of SCBWP are based on model calculations presented by Collins \&
Sonneborn (1977).  They compared the equivalent width ratio of
\ionI{He}4471\ to \ionII{Mg}4481\ to the model calculations to derive
\vsini. This procedure is not easy to compare to what we have done.


However, if we use the line-widths for \ionII{Mg}4481\ as measured by
SCBWP and calculate \vsini\ according to the modern models of CTC the
discrepancies largely disappear and are no longer systematic. The new
\vsini\ values were derived directly from the \ionII{Mg}4481\
line-widths according to the procedure described in Collins
\& Truax (1995) after correction for the intrinsic and instrumental
broadening of the line. Among the sharp-lined stars there are also two
giants for which SCBWP used main-sequence models to derive
\vsini, which leads to underestimated values.
Therefore, we conclude that our derived values of \vsini\ are on the
same velocity scale as defined by SCBWP.

\titlea{The Distribution of Rotational Velocities in Sco OB2}
In this section we discuss the distribution of projected and true
rotational velocities in Sco OB2. We concentrate on the known members
of the association. Of the 156 stars we observed 88 are known members
of the association (counting the double-lined spectroscopic binary
HD 140008 as two objects). In order to increase the sample size we
added \vsini\ values from literature for an additional 48 members
(counting HD 145501 twice). The membership of these stars was taken
from de Geus \ea (1989) and Bertiau (1958). The \vsini\ values were
taken from Slettebak (1968) and Uesugi \& Fukuda (1981). Their values
are on the old velocity scale defined by the studies of Slettebak
(1954, 1955, 1956, 1966a, 1966b, 1968) and Slettebak \& Howard (1955),
whereas our \vsini\ values are on the velocity scale defined by
SCBWP. In order to take the differences into account we corrected the
literature values downwards by 15\% and 5\% for B-type stars and
A-type stars, respectively, as recommended by SCBWP. In Table 4 we
give the resulting values of \vsini\ for the literature sample. The
columns have the same meaning as in Table 2. The stars HD 120908 and
HD 138690 are members of UCL, all others are members of US. Note that
HD 147889 is a doubtful member according to Bertiau (1958) and de Geus
\ea (1989) do not list this star as a member. However, including or
excluding this star from the analysis does not influence the results.

\begtabfull
\tabcap{4} {Projected rotational velocities for the literature sample.}

\halign{\strut#\hfil & \quad#\hfil & \hfil#
            & \quad\hfil# & \quad#\hfil & \hfil#\hfil & #\hfil \cr
\noalign{\smallskip\hrule\vskip2pt}
\hfil HD & \hfil MK & \vsini\hfil & \vbreak\hfil &
	\hfil$\vsini\over\vbreak$ & Mem. &\hfil Dupl. \cr
\noalign{\vskip2pt\hrule\vskip2pt}
120908 & B5III       &  85 & 308 & $0.276$ & M & \omit \cr
138690 & B2IV        & 229 & 412 & $0.557$ & M & SB1 \cr
138485 & B2Vn        & 212 & 459 & $0.463$ & M & RV  \cr
138764 & B6IV        &  17 & 342 & $0.050$ & M & RV \cr
139094 & B8IV/V      & 153 & 363 & $0.421$ & M & \omit \cr
\noalign{\vskip3pt}
139486 & B9V         & 212 & 350 & $0.607$ & M & RV \cr
141404 & B9.5IV      & 170 & 350 & $0.486$ & M & \omit \cr
141774 & B9V         & 136 & 350 & $0.389$ & M & RV \cr
142165 & B5V         & 204 & 388 & $0.526$ & M & SB1 \cr
142250 & B7V         &  42 & 373 & $0.114$ & M & RV \cr
\noalign{\vskip3pt}
142315 & B9V         & 255 & 350 & $0.729$ & M & SB1 \cr
142378 & B2/B3V      & 204 & 411 & $0.496$ & M & RV \cr
142884 & B8/B9III    & 170 & 283 & $0.601$ & m & CON \cr
143567 & B9V         & 153 & 350 & $0.437$ & M & RV \cr
143600 & B9.5V       & 255 & 350 & $0.729$ & M & RV \cr
\noalign{\vskip3pt}
143699 & B6III/IV    & 144 & 303 & $0.477$ & M & RV \cr
144334 & B8V         &  38 & 363 & $0.105$ & M & \omit \cr
144661 & B8IV/V      &  85 & 363 & $0.234$ & M & RV \cr
145102 & B9VpSi      &  34 & 350 & $0.097$ & M & RV \cr
145353 & B9V         & 187 & 350 & $0.534$ & M & \omit \cr
\noalign{\vskip3pt}
145501(1) & B8V+B9VpSi  &  59 & 357 & $0.167$ & m & RV \cr
145501(2) & B8V+B9VpSi  &  63 & 357 & $0.179$ & m & RV \cr
145519 & B9Vn        & 255 & 350 & $0.729$ & M & SB1 \cr
145554 & B9V         & 153 & 350 & $0.437$ & M & CON \cr
145631 & B9V         & 153 & 350 & $0.437$ & M & CON \cr
\noalign{\vskip3pt}
145792 & B6IV        &  21 & 342 & $0.062$ & M & CON \cr
146001 & B8V         & 170 & 363 & $0.468$ & M & CON \cr
146029 & B9V         & 212 & 350 & $0.607$ & M & CON \cr
146284 & B9III/IV    & 170 & 278 & $0.612$ & M & \omit \cr
146285 & B8V         & 136 & 363 & $0.375$ & M & RV \cr
\noalign{\vskip3pt}
146416 & B9V         & 255 & 350 & $0.729$ & M & CON \cr
146706 & B9V         & 229 & 350 & $0.656$ & m & \omit \cr
146998 & A2pSr       &  23 & 350 & $0.068$ & m & \omit \cr
147009 & B9.5V       & 136 & 350 & $0.389$ & M & CON \cr
147010 & B9II/III    &  21 & 278 & $0.076$ & m & RV \cr
\noalign{\vskip3pt}
147196 & B6/B7Vn     & 297 & 377 & $0.789$ & M & \omit \cr
147701 & B5III       &  68 & 308 & $0.221$ & m & \omit \cr
147703 & B9Vn        & 238 & 350 & $0.680$ & m & \omit \cr
147888 & B3/B4V      & 153 & 406 & $0.377$ & M & CON \cr
147889 & B2III/IV    &  85 & 364 & $0.234$ & M & RV \cr
\noalign{\vskip3pt}
147890 & B9.5pSi     &  25 & 350 & $0.073$ & M & RV \cr
147932 & B5V         & 153 & 388 & $0.394$ & m & CON \cr
147955 & B9V         & 238 & 350 & $0.680$ & m & \omit \cr
148199 & B9VSi       &  17 & 350 & $0.049$ & m & CON \cr
148579 & B9V         & 127 & 350 & $0.364$ & M & RV \cr
\noalign{\vskip3pt}
148594 & B8Vnn       & 255 & 363 & $0.702$ & M & CON \cr
148605 & B2V         & 195 & 459 & $0.426$ & M & CON \cr
148860 & B9.5V       & 255 & 350 & $0.729$ & m & \omit \cr
\noalign{\vskip2pt\hrule}}


\endtab

Table 5 shows the distribution among spectral type for the members of
the subgroups of Sco OB2. The numbers in parentheses indicate the
stars for which \vsini\ was obtained from literature. Note that the
sample observed by us is very biased towards spectral types B0--B3
(see Sect.~2). If one assumes the initial mass function given in de
Geus (1992) for Sco OB2, the numbers indicate that our observations
are far from complete. Note also that of the 48 \vsini\ values from
literature 46 are for stars in US and of these 33 are of spectral type
B7--B9. All this means that caution should be exercised when comparing
distributions of \vsini\ for the various subgroups.

\begtabfull
\tabcap{5} {Distribution of spectral types for the members of Sco OB2.}
\halign{\strut#\hfil & \quad\hfil#\hfil & \quad#\hfil
            & \quad#\hfil & \quad#\hfil & \quad#\hfil \cr
\noalign{\smallskip\hrule\vskip3pt}
Subgroup\hfil & O & B0--B3 & B4--B6 & B7--B9 & A \cr
\noalign{\vskip3pt\hrule\vskip3pt}
 US & 1 & 25 (5) & 3 (7) & 3 (33) & 1 (1) \cr
UCL & 0 & 18 (1) & 6 (1) & 5      & 0     \cr
LCC & 0 & 18     & 6     & 2      & 0     \cr
\noalign{\vskip3pt\hrule}}
\endtab

\titleb{The \vsini\ Distributions in the Sco OB2
Subgroups and the Binary Population} Figures 9 and 10 show the
distributions of \vsini\ and $\vsini/\vbreak$ for Sco OB2 and each of
the subgroups for which membership studies have been done. In both
figures we indicate the distributions of single stars and binaries
(spectroscopic binaries and radial velocity variables) separately,
with solid and dotted lines respectively. Both figures suggest that
amongst the binaries there is a higher proportion of slow rotators. We
tested this by performing a robust-rank-order test on the members of
Sco OB2 that we have observed and for which we have good information
on duplicity (see Table 2). This is a non-parametric statistical test
that is designed to test for differences in the median of two samples
(see e.g., Siegel \& Castellan 1988). Nothing is assumed a priori
about the underlying distributions. For the
\vsini\ distributions this test confirms at a confidence level of more
than $99.99$\% that the single stars have a higher median \vsini. For
the $\vsini/\vbreak$ distributions the same result is obtained. This
implies that the bulk of the single stars has a higher projected
rotational velocity. We find the same result if we use all the members
of Sco OB2 (assuming that stars for which we have no duplicity
information are single).

\begfig 8.8cm
\vskip -8.8cm
\psfig{figure=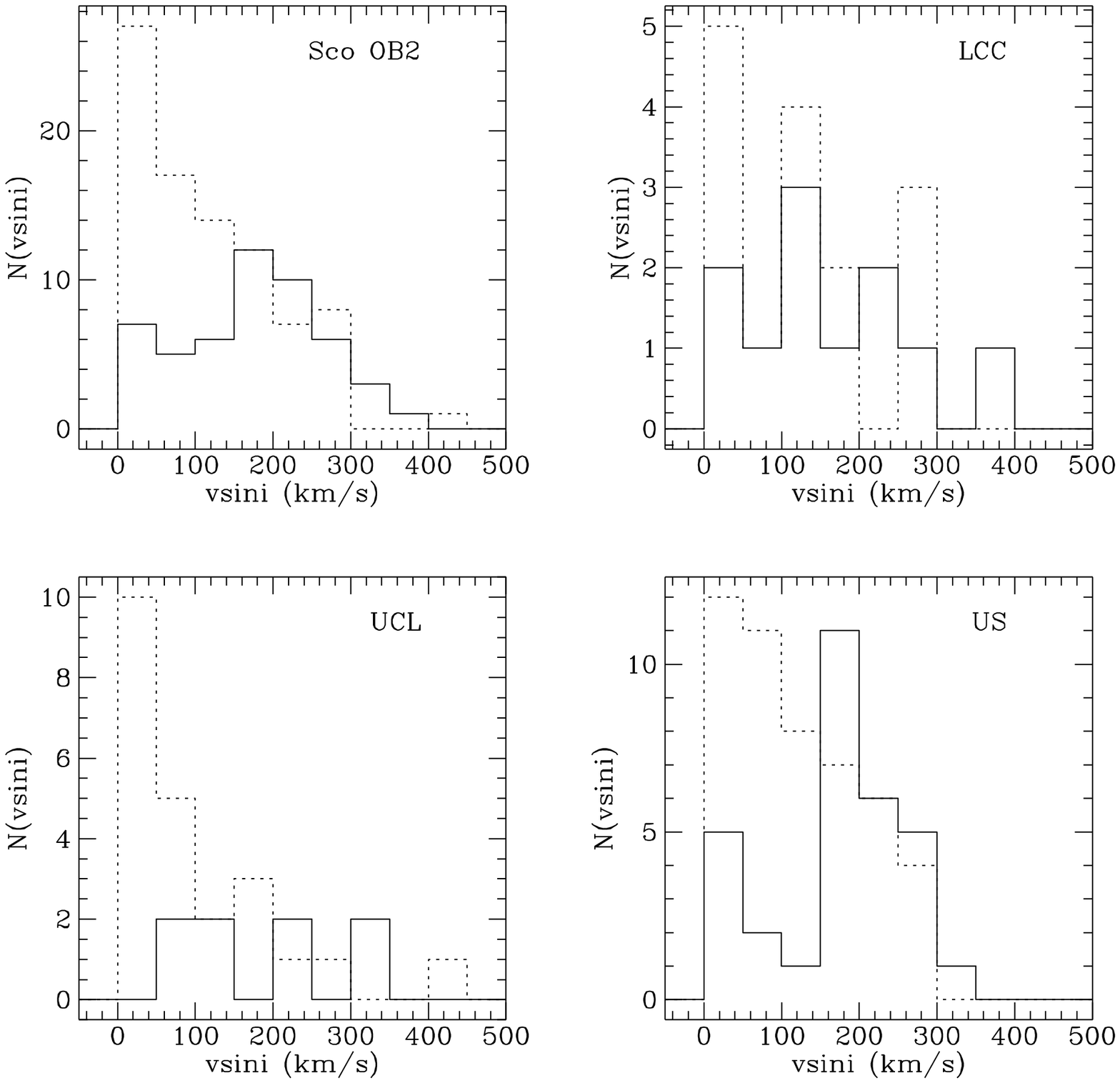,height=8.8cm}
\figure{9} {The distribution of \vsini\ values for the
members of Sco OB2. The distribution for the whole association is
shown as well as for each subgroup separately. The solid lines show
the distribution for the single stars and the dotted lines show the
distribution for the binaries (radial velocity variables and known
spectroscopic binaries).}
\endfig

\begfig 8.8cm
\vskip -8.8cm
\psfig{figure=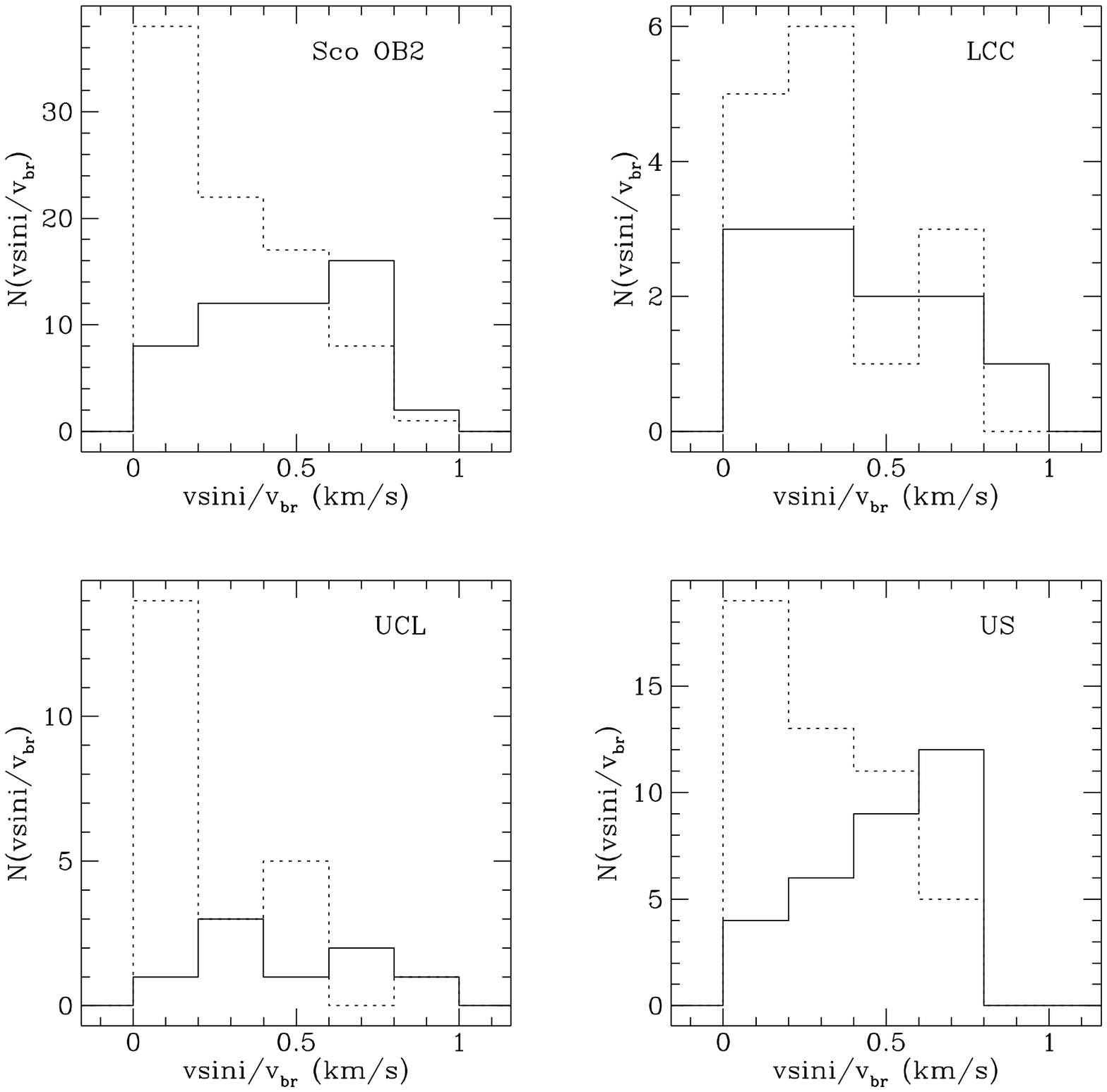,height=8.8cm}
\figure{10} {Same as Fig.~9 for the distribution of
$\vsini/\vbreak$.}
\endfig

The finding that binaries show lower rotational velocities can be
interpreted as a consequence of the deposition of part of the total
angular momentum in the orbit of the system at formation or as a
consequence of tidal braking in very close binaries. Levato
\ea (1987) have shown, using the \vsini\ data of Slettebak (1968),
that there is a correlation between the frequency of duplicity and the
average \vsini\ for the subgroups US and UCL. This correlation is in
the sense that stars in UCL rotate slower on average and that the
frequency of duplicity is higher in that subgroup. Based on the
duplicity data in Tables 2 and 4 we also find a higher proportion of
binaries in UCL than in US. However the distributions of rotational
velocities are not significantly different for these two subgroups.

If one looks at the data in more detail it turns out that most of the
single stars at high \vsini\ values are stars added from the
literature in the spectral type range B7--B9, whereas our sample
contains mainly B0--B6 stars (see Sect.~2). Most of the literature
data (46 out of 48 stars) was added to US and indeed the difference
between the single and binary stars is most pronounced for this
subgroup (see Figs.~9 and 10). In US the extra
literature data consist of 28 stars for which $\vsini\geq150\kms$. Of
these stars 15 are single stars in the B7--B9 range. The sample
observed by us contains, in contrast, only one single B7--B9 star in
the same \vsini\ range. To check whether the stars in the B7--B9 range
rotate faster we divided the sample in B0--B6 and B7--B9 single
stars. Each of these samples was then divided into stars for which
$\vsini\geq150\kms$ and $\vsini<150\kms$. Subsequently, a $2\times2$
contingency table analysis was done on the data with the Fisher exact
probability test (see e.g., Siegel \& Castellan 1988). The test
confirms at a confidence level of more than $99.99$\% that the B7--B9
single star sample contains more objects with $\vsini\geq150\kms$. If
one does the same test for $\vsini/\vbreak$ with the division of the
samples at $0.4$ then the same result is found at the $99.98$\%
confidence level. The value $0.4$ follows from the fact the all B7--B9
stars with $\vsini\geq150\kms$ have $\vsini/\vbreak\geq0.4$.  Thus, it
could be that late B-stars exhibit intrinsically faster
rotation. However, this finding is based on inhomogeneous \vsini\
data. To really interpret the \vsini\ data, a more complete coverage
of the spectral types is needed in all subgroups of Sco OB2 and a
complete census of the binary population as a function of spectral
type is also required.

Finally, we remark that there are no significant differences between
the subgroups of Sco OB2 for the median value and the overall
distribution of \vsini\ or $\vsini/\vbreak$.

\titleb{Tests of Various Hypotheses about the Distribution of
Intrinsic Rotational Velocities} We now test various hypotheses about
the distribution of intrinsic rotational velocities against the
data. We wish to investigate whether any constraints can be placed on
the distributions of true rotational velocities and/or the orientation
of rotation axes. This will lead to constraints on the origin of
angular momentum in stars and on the star formation process. So far
the only models for \vsini\ distributions that provide detailed
predictions are the ones that assume a random selection process. The
magnitude of the rotational velocity and the inclination angle of the
rotation axis are selected separately and randomly according to a
certain prescription. These models do not take into account the
physically more likely possibility that a star is formed with a
certain angular momentum. The role of stellar mass and the star
formation process is also ignored.

We assume in all cases that the rotation axes are oriented randomly in
space. This means that the distribution of inclination angles $i$ is
given by:
$$
	f(i)\, di=\sin i\, di \,.	\eqno\autnum
$$
In what follows we define \veq\ to be the true equatorial rotational
velocity of the star, $\eta=\veq/\vbreak$ and
$y=\vsini/\vbreak$. Again, \vbreak\ is the true equatorial velocity of
the star at break-up, and $w$ is the fraction of the corresponding
critical angular velocity. Note that in case the Roche model is
assumed for the rotating star, \vsini\ is not equal to $\vbreak\times
w\times\sin i$ and that it makes a difference whether $\eta$ or $w$ is
chosen randomly to generate the distribution of $y$. In the Roche
model the star rotates uniformly and its surface follows an
equipotential surface. The potential is generated by a point mass at
the centre of the star. The equation describing the surface is:
$$
	\Phi = {{GM}\over R} + {1\over2}\omega^2R^2\sin^2\theta = 
		{{GM}\over R_{\rm p}}\,.		\eqno\autnum
$$
$\Phi$ is the total potential consisting of a gravitational and a
centrifugal term. The mass of the star is $M$ and the distance from
the surface to the centre is given by $R$. The polar radius is $R_{\rm
p}$, which remains constant, and $\theta$ is the latitude on the
surface of the star as measured from the pole. The critical angular
velocity follows by setting $\vec g=-\vec\nabla\Phi=0$ and is given
by: $\omega^2_{\rm c}=8GM/27R^3_{\rm p}$. With $w=\omega/\omega_{\rm
c}$ and setting $x=R/R_{\rm p}$ Eq.~(2) becomes:
$$
	{1\over x} + {4\over27}w^2x^2\sin^2\theta = 1\,. \eqno\autnum
$$
Now, from the definitions of $\omega_{\rm c}$, $x$ and $w$ it follows
that $\eta={2\over3}wx$, where $x$ is to be evaluated at the
equator. This value of $x$ follows by setting $\theta=\pi/2$ in
Eq.~(3) and solving this equation for $x$ for a given $w$. The
equation can be solved analytically and it follows that:
$$
	\eta=-2\cos\left({{\arccos w-2\pi}\over 3}\right)\,.  \eqno\autnum
$$
It is this transformation between $w$ and $\eta$ that one has to take
into account if the distribution of $y$ is generated by picking $w$
randomly.

We tested the following models for the distribution of intrinsic
rotational velocities of the stars:
\medskip
\item{1.} The value of $w$ is distributed homogeneously on $[0,1]$.
\item{2.} The value of $\eta$ is distributed homogeneously on $[0,1]$.
\item{3.} The distribution of \veq\ is given by a Maxwellian.
\item{4.} The distribution of $w$ is given by a truncated Maxwellian.
\medskip

\noindent
All the hypotheses about the distribution of \veq\ are variations on
the assumption of a completely random distribution of rotational
velocities. So the question is whether one can exclude any form of
total randomness in the \veq\ or $w$ distribution.  If all these
models are excluded by the data then either one has to relax the
assumption of random orientation of rotation axes or a non-random
process for generating the distribution of rotational velocities has
to be assumed. We find no correlation in our data between \vsini\ and
Galactic longitude, which is consistent with the assumption of random
orientation of rotation axes.

For the four models above one can derive an expression for the
distribution of $y$ by using the formula given by Chandrasekhar \&
M\"unch (1950):
$$
	\phi(\vsini)=\vsini\int^\infty_\vsini
		{{f(\veq)}\over{\veq\sqrt{\veq^2-(v\sin i)^2}}}
		\, d\veq\, , 	\eqno\autnum
$$
where $\phi(\vsini)$ is the distribution function of $\vsini$ and
$f(\veq)$ that of \veq. If there is a break-up velocity, then
Eq.~(5) becomes (in terms of $y$ and $\eta$):
$$
	\phi(y)=y\int^1_y{{f(\eta)}\over{\eta\sqrt{\eta^2-y^2}}}
		\, d\eta\, , 	\eqno\autnum
$$
The form of $\phi(y)$ for model 2 is derived in Chandrasekhar \&
M\"unch (1950, their Eq.~(26)). To derive $\phi(y)$ for model 1 one
should use
$$
	f(\eta)=f(w)\vert{{dw}\over{d\eta}}\vert\, ,	\eqno\autnum
$$
with:
$$	
	\vert{{dw}\over{d\eta}}\vert={3\over2}(1-\eta^2)\, .	\eqno\autnum
$$
Deutsch (1970) showed, using arguments from statistical mechanics,
that if $\vec\omega$ is distributed isotropically with mutually
independent Cartesian components the distribution of $\omega$ is given
by a Maxwellian. The form of the distribution function is:
$$
	f(z)={4\over{\sqrt\pi}}z^2{\rm e}^{-z^2}\, , \eqno\autnum
$$
where $z=j\omega$ and $j$ is a parameter with the dimensions
$\omega^{-1}$. It follows that the distribution of \vsini\ is then given
by:
$$
	\phi(u)=2u{\rm e}^{-u^2}\, ,	\eqno\autnum
$$
where $u$ is $j\vsini$ and $1/j$ is the root mean square value of
\vsini\ which is the same as the modal value of \veq\ for a
Maxwellian (see Deutsch 1970). This is the distribution of \vsini\ for
model 3. Note that it is assumed that there is no break-up velocity in
this case. To derive the distribution for model 4 one should take into
account that $\omega\leq\omega_{\rm c}$. We chose to model this simply
by truncating the Maxwellian at $\omega_{\rm c}$ and normalizing the
resulting distribution. In terms of $\eta$ one can find $\phi(y)$ by
using Eqs.~(4), (6) and (7). The distribution of $w$ is now given by:
$$
	f(w)={{{4\over{\sqrt\pi}}w^2{\rm e}^{-w^2}}\over
		{{\rm erf}(1)-{2\over{\sqrt\pi}}{\rm e}^{-1}}}\, , \eqno\autnum
$$
\noindent
where erf is the error function.

\begfigside 6cm 12cm
\hskip -12.5cm
\psfig{figure=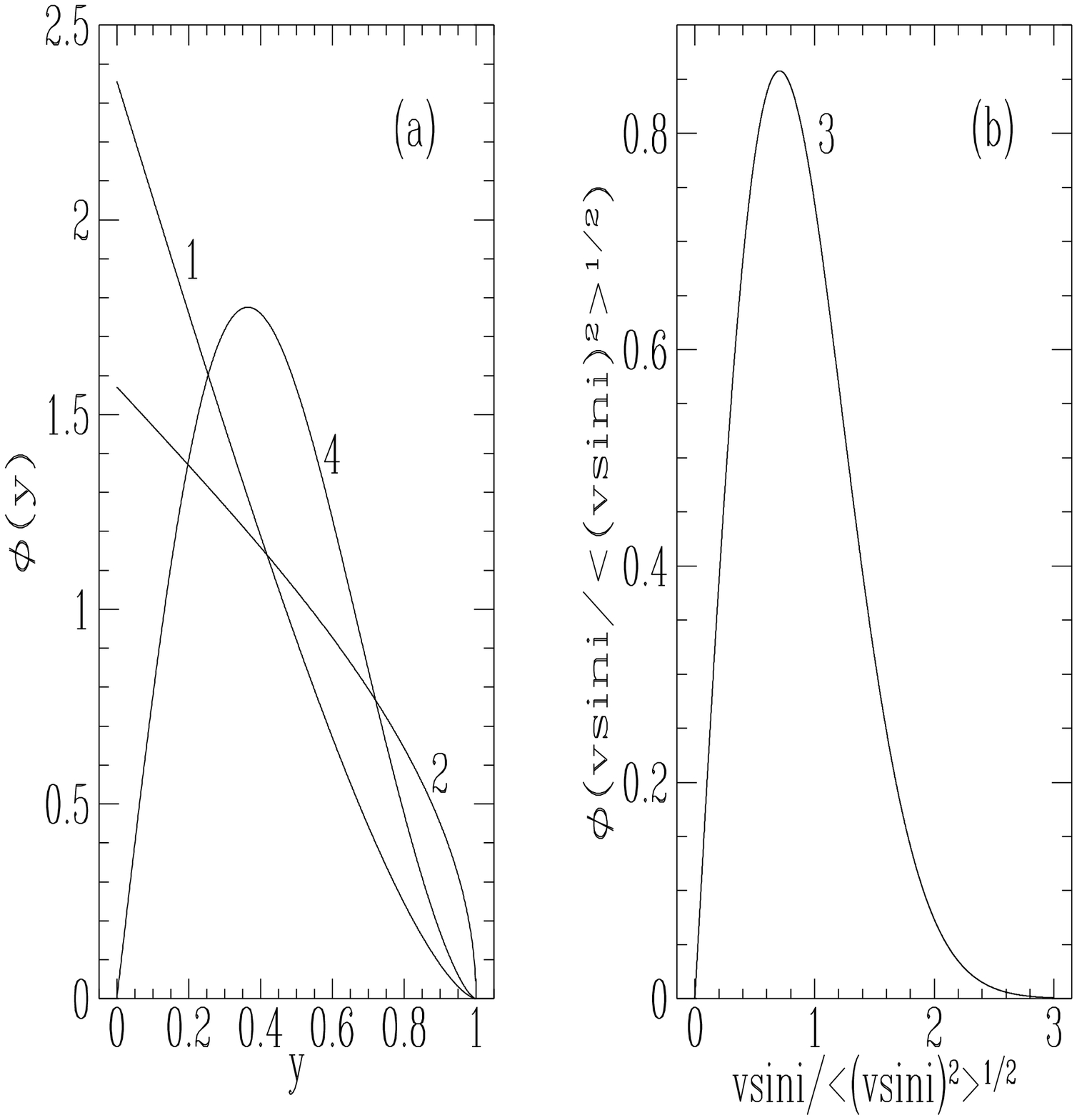,height=6cm,width=12cm}\hfil
\figure{11} {(a) The distribution functions of $y=\vsini/\vbreak$
for models 1, 2 and 4. (b) The distribution function of \vsini\ for
model 3. The models are indicated by their corresponding numbers.}
\endfig

The distribution functions for $y$ for models 1, 2 and 4 are listed in
Table 6 (model 3 being given by Eq.~(10)). For model 4 the
integral in Eq.~(6) has to be evaluated numerically. All the
$\phi(y)$ are shown in Fig.~11. Note that models 3 and 4 show a
peak in the distribution at intermediate values of \vsini\ and
$\vsini/\vbreak$ whereas models 1 and 2 predict a large number of slow
rotators.

\begtabfull
\tabcap{6} {Distribution functions for $y=\vsini/\vbreak$.}
\halign{\strut\hfil#\hfil & \hfil $\displaystyle{#}$ \hfil 
			  & \hfil $\displaystyle{#}$ \hfil \cr
\noalign{\smallskip\hrule\vskip3pt}
Model & f(\eta) & \phi(y) \cr
\noalign{\vskip3pt\hrule\vskip3pt}
1 & {3\over2}(1-\eta^2) & -{3\over2}y\sqrt{1-y^2}+{3\over2}\arccos y\cr
\noalign{\bigskip\bigskip}
2  & 1 & \arccos y \cr
\noalign{\bigskip\bigskip}
4  & {3\over2}(1-\eta^2){{4\over\sqrt\pi}w^2{\rm e}^{-w^2}\over
	{\rm erf}(1)-{2\over\sqrt\pi}{\rm e}^{-1}}\,; & 
	y\int_y^1 {f(\eta)\over\eta\sqrt{\eta^2-y^2}}\, d\eta \cr
\noalign{\smallskip}
\omit & w=\cos(\arccos({-\eta\over2})) & \omit \cr
\noalign{\vskip3pt\hrule}}
\endtab


The models were compared to the data by calculating the
Kolmogorov-Smirnoff statistic $D$ and the probability $P$ that $D$
exceeds the observed value under the hypothesis that the data were
drawn from the model distribution (see e.g., Press \ea 1992). A model
was rejected if $P<0.05$. As argued in the previous section, the
B7--B9 stars may be a group of stars that rotate intrinsically faster
than B0--B6 stars. Our data consist primarily of B0--B6 stars and
adding literature data on later types only to US may introduce
artifacts in the \vsini\ distribution that are the result of our
sample selection. Thus we decided to compare the model distributions
only to the B0--B6 stars.

We did the KS-test for Sco OB2 as a whole and for each subgroup. The
data were also divided in single stars and binaries. The binaries
include the spectroscopic binaries as well as the radial velocity
variables. The comparison of the models to the data was focused on the
members of Sco OB2. We are interested here in the distribution of
rotational velocities in a group of physically related stars. The
results are shown in Table 7. In this table a ``$+$'' indicates that
the model was not rejected and a ``$-$'' indicates that the model was
rejected.

\begtabfull
\tabcap{7} {Results of comparing model distributions to the data
(B0--B6 members). A ``-'' indicates that the model is rejected. A ``+''
indicates that the model is not rejected. The number of stars in each
data set is indicated in parentheses.}
\offinterlineskip
\halign{&\vrule# & \strut\hskip6pt\hfil#\hfil
	&\vrule# & \strut\hskip6pt\hfil#\hfil
	         & \strut\quad\hfil#\hfil
	         & \strut\quad\hfil#\hfil
        &\vrule# & \strut\hskip6pt\hfil#\hfil
	         & \strut\quad\hfil#\hfil
	         & \strut\quad\hfil#\hfil \cr
\noalign{\smallskip\hrule}
height2pt & \omit && \omit & \omit & \omit && \omit & \omit & \omit & \cr
& \omit && \multispan3 \hfil Sco OB2\hfil && \multispan3 \hfil US\hfil & \cr
height3pt & \omit && \omit & \omit & \omit && \omit & \omit & \omit & \cr
& Model && All & Single & Binary && All & Single & Binary & \cr
height3pt & \omit && \omit & \omit & \omit && \omit & \omit & \omit & \cr
& \omit && (90) & (23) & (67) && (40) & (9) & (31) & \cr
height2pt & \omit && \omit & \omit & \omit && \omit & \omit & \omit & \cr
\noalign{\hrule}
height2pt & \omit && \omit & \omit & \omit && \omit & \omit & \omit & \cr
& 1 && + & + & + && + & + & + & \cr
& 2 && - & + & - && + & + & - & \cr
& 3 && - & + & - && + & + & + & \cr
& 4 && - & + & - && - & + & - & \cr
height2pt & \omit && \omit & \omit & \omit && \omit & \omit & \omit &\cr
\noalign{\hrule}
height2pt & \omit && \omit & \omit & \omit && \omit & \omit & \omit & \cr
& \omit && \multispan3 \hfil UCL\hfil && \multispan3 \hfil LCC\hfil & \cr
height3pt & \omit && \omit & \omit & \omit && \omit & \omit & \omit & \cr
& Model && All & Single & Binary && All & Single & Binary & \cr
height3pt & \omit && \omit & \omit & \omit && \omit & \omit & \omit & \cr
& \omit && (26) & (5) & (21) && (24) & (9) & (15) & \cr
height2pt & \omit && \omit & \omit & \omit && \omit & \omit & \omit & \cr
\noalign{\hrule}
height2pt & \omit && \omit & \omit & \omit && \omit & \omit & \omit & \cr
& 1 && + & + & + && + & + & + & \cr
& 2 && - & + & - && + & + & + & \cr
& 3 && - & + & - && + & + & + & \cr
& 4 && - & + & - && - & + & + & \cr
height2pt & \omit && \omit & \omit & \omit && \omit & \omit & \omit &\cr
\noalign{\hrule}}
\endtab

The number of measurements for each subgroup is rather small (40 stars
in US, 26 in UCL and 24 in LCC), which makes it difficult to
distinguish between models. The data for the single stars are
compatible with all models. This is probably the result of the limited
amount of data. If we take only the data for US into account and
include the B7--B9 stars we find that only model 4 is consistent with
the data. It is not warranted, however, to conclude that the formation
of single stars leads to an isotropic distribution of angular momentum
vectors. We have shown in the previous section that the B7--B9 stars
rotate faster than B0--B6 stars. So there is an effect with stellar
mass that model 4 does not take into account.

For the binary stars it seems that only model 1 is consistent with the
data. What does it mean if we assume that this model actually
describes the data? Do components of binary stars acquire a random
fraction of their critical angular velocity or is model 1 the only
model that predicts enough slow rotators? Remember that binary
components show intrinsically slower rotation than the single
stars. This may be due to the redistribution of angular momentum into
orbital angular momentum.

If one aims at putting real constraints on the star formation process
by looking at the distribution of rotational velocities, then models
are needed that take into account the details of this
process. How does the combination of stellar mass, angular momentum
and interior structure translate into an observed rotational velocity?
So far no detailed predictions of this kind have been
made. Mouschovias (1983) does give an estimate of the equatorial
velocity of the star as it arrives on the zero-age main-sequence. His
prediction follows from a consideration of magnetic braking in the
early stages of star formation. As soon as the protostellar
condensation is dense enough to decouple from the surrounding magnetic
field its angular momentum is basically determined. It then follows
that:
$$
	\veq\propto\rho^{2/3}RM_{\rm C}\,,  \eqno\autnum
$$
where $R$ and $\rho$ are the mean density and radius of the protostar
and $M_{\rm C}$ is the mass of the cloud-core from which the protostar
formed. The factor $\rho^{2/3}R$ increases going from O to F0
stars. If one assumes that the mass of the cloud-core scales with the
mass of the star that forms from it, this equation predicts a maximum
in the rotational velocities somewhere between O and F stars. This is
consistent with the observed distribution of velocities for field
stars (e.g., Wolff \ea\ 1982), and with our observed distribution of
rotational velocities as a function of spectral type.

Another model for the origin of angular momentum in stars that takes
the star formation process into account is given by Wolff
\ea (1982). They find a large number of slow rotators in their data
that cannot be explained by the presence of binaries. Therefore they
propose that all protostars start out in a state of low angular
momentum. Subsequently, a number of them acquire higher angular
momenta through gravitational interactions with other protostars or
protostellar condensations in a forming cluster. This model then
predicts a concentration of rapid rotators in the centres of clusters
and associations. We cannot see this clearly in our data and in our
case the presence of binaries can account for the excess of slow
rotators with respect to a Maxwellian.

To go further more detailed predictions of the distribution of
rotational velocities are needed that take the star formation process
into account. Observationally, more data are required (\vsini\ as well
as duplicity information) for the members of Sco OB2 of spectral types
later than B6. This will facilitate a study of \vsini\ as a function of
mass and of the relation to duplicity.

\titlea{Rectifying the \vsini\ Distribution}
In the previous section we compared model \veq\ and $\veq/\vbreak$
distributions to the observed \vsini\ distribution in Sco OB2. In this
section we present the distributions of \veq\ and $\veq/\vbreak$ that
follow directly from the rectification of the observed
\vsini\ distribution under the (sole) assumption of randomly oriented
rotation axes. Two techniques are used to do so; Lucy-iteration (Lucy,
1974) and the technique presented by Bernacca (1970).

\begfig 8.8cm
\vskip -8.8cm
\psfig{figure=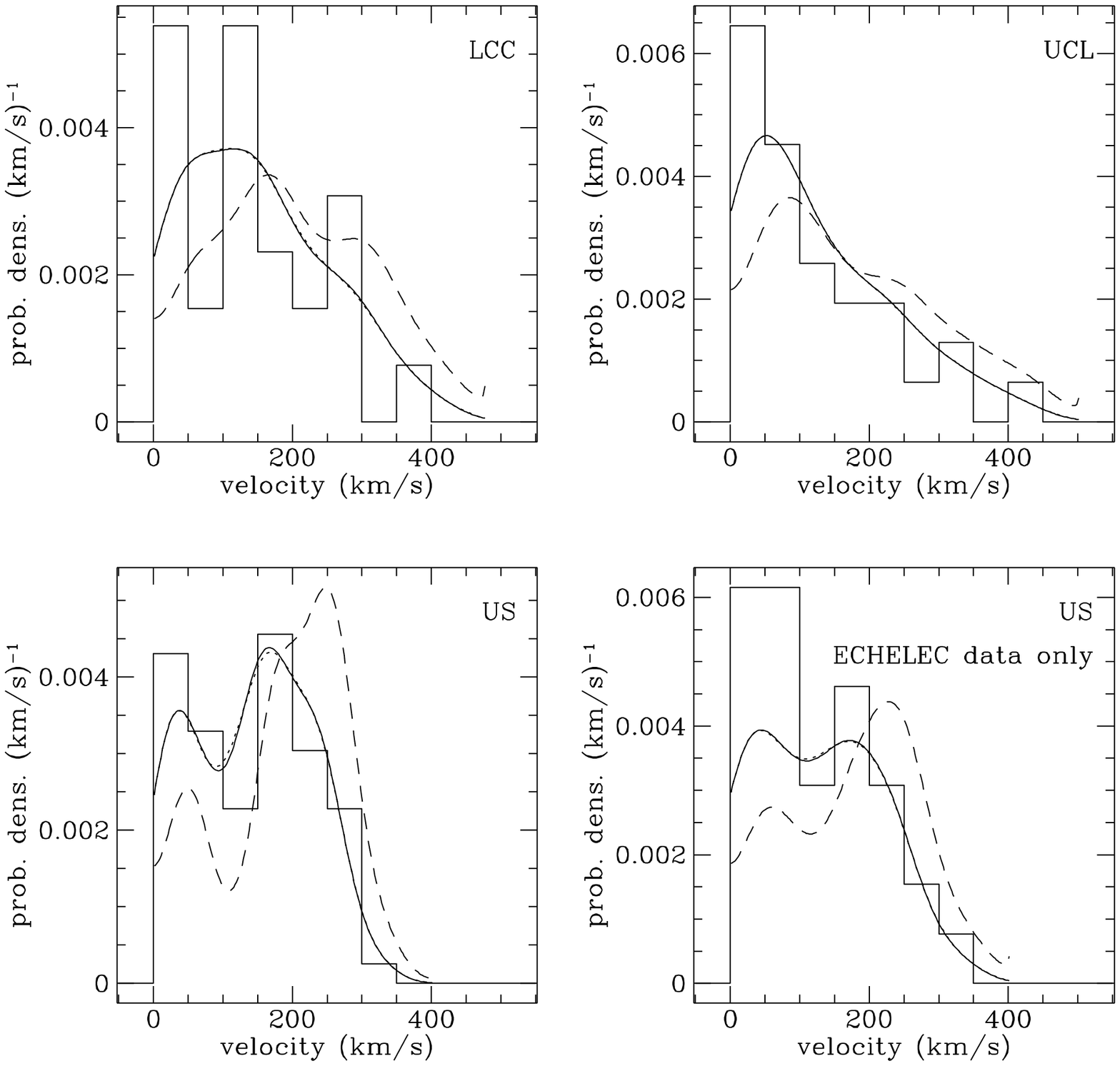,height=8.8cm}
\figure{12} {Results of applying the Lucy iterative technique
for the rectification of \vsini\ distributions. The number of
iterations used was 8, 9, 11 and 9 for LCC, UCL, US and the ECHELEC
data in US, respectively. The data histograms are shown for
comparison. The solid lines indicate $\tilde\varphi(\vsini)$, the
dashed lines are for $\psi(\veq)$ and the short-dashed lines are for
$\varphi(\vsini)$.}
\endfig

For both techniques it is necessary to have a good estimate of the
observed probability density function (PDF). We used a so-called
kernel estimator. The PDF is represented as a sum of kernel functions,
each corresponding to a single data point. The kernel function is
normalized to unity and it is characterized by a width $h$. The width
depends on the observed distribution of the data. A rule of thumb for
calculating $h$ is given in Vio \ea (1994). The kernel estimator
provides a smooth estimate of the PDF. We used Gaussian kernels, in
which case $h$ corresponds to the standard deviation.

For both rectification techniques the interval for the kernel-estimate
of the PDF was restricted to $\vsini\geq0\kms$. For the Lucy-iteration
the upper limit of the interval was set by increasing the data
interval in small steps until the cumulative distribution increased by
no more than $0.01$\%. In the technique given by Bernacca (1970) one
assumes that there is an upper limit to \vsini. We used the break-up
velocity of the stars as a function of spectral type and rectified the
$\vsini/\vbreak$ distribution. So the PDF was restricted between 0 and
1. In both cases the kernel-estimator formally has tails outside the
intervals mentioned above, and we decided to simply re-normalize the
PDF on the restricted intervals.

For the iterative technique given by Lucy (1974) we used the optimum
stopping criterion described in Lucy (1994). In his paper two
quantities are defined:
$$
	H=\sum_{i=1}^N \tilde\varphi_i\ln\varphi_i	\eqno\autnum
$$
and
$$
	S=-\sum_{i=1}^N \psi_i\ln\psi_i\,, \eqno\autnum
$$
where $H$ is $N^{-1}\times {\rm the}$ log-likelihood of the data
vector $\tilde\varphi$ (this vector being the estimate of the PDF
described above), with $\varphi$ the data vector that one derives from
the rectified distribution $\psi$. $S$ is a function that decreases as
the solution $\psi$ increases in complexity (Note that we took $\chi_j$
constant in Eq.~(6) of Lucy 1994). The solution $\psi$ will describe a
trajectory in the $(H,S)$ plane during iteration and one can define
the curvature of this trajectory as:
$$
	\kappa={\left|\dot S\ddot H-\dot H\ddot S\right|\over
	       \left(\dot S^2-\dot H^2\right)}\,. \eqno\autnum
$$
It turns out that the iterations should be stopped when $\kappa$
passes through a minimum.

\begfigside 6cm 12cm
\hskip -12.5cm
\psfig{figure=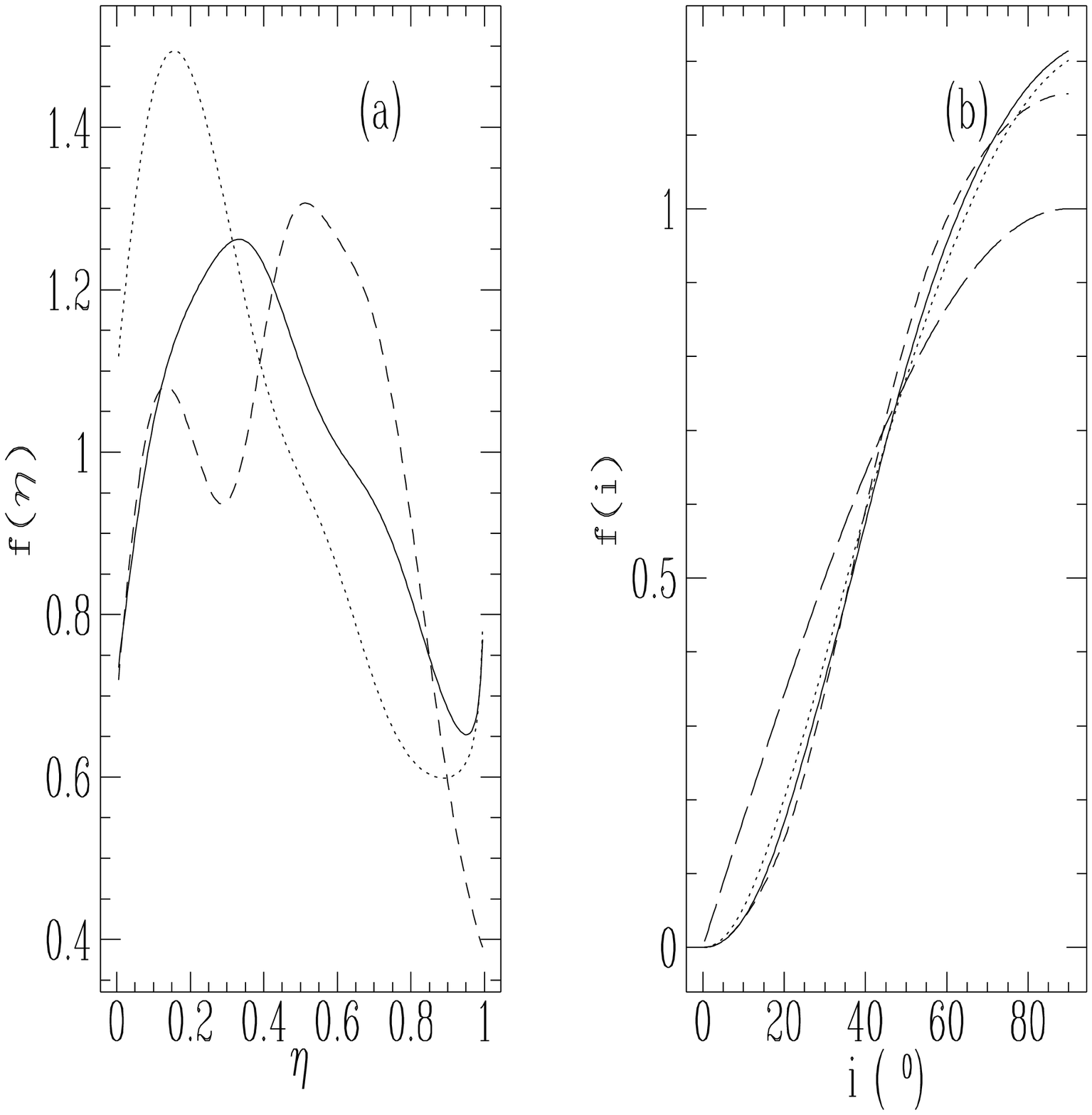,height=6cm,width=12cm}\hfil
\figure{13} {(a) shows the distributions of
$\eta=\veq/\vbreak$ for the subgroups of Sco OB2 derived with the
technique of Bernacca (1970). The solid line indicates the result for
LCC, the short-dashed line that for UCL and the dashed line indicates
the result for US. The up-turns in the distributions for LCC and UCL
at $\eta=1$ are probably not real but due to the singularity in the
integral in Eq.~(18). (b) shows the corresponding distributions of
$i$. In this figure the long-dashed line indicates the distribution
$f(i)=\sin i$.}
\endfig

Figure 12 shows the results of rectifying the observed \vsini\
distributions for the members of each subgroup of Sco OB2. For US we
also show the result of rectification if we use only our own data. In
all figures the data histogram is shown. The solid line is the
estimate of the observed PDF $\tilde\varphi(\vsini)$, the dashed lined
is the rectified distribution $\psi(\veq)$ and the short-dashed line
is the PDF $\varphi(\vsini)$ that follows from $\psi$.

Figure 12 suggests that the peak in the \veq\ distribution lies at
increasingly higher values going from UCL to LCC to US. This result is
the same even if only our own data for US are used. Whether this
effect is due to a decrease in binary frequency per subgroup is hard
to tell. The data are incomplete for the B7--B9 stars for UCL and
LCC. If one takes only B0--B6 stars into account and only our sample
the binary frequency increases from LCC to UCL to US, opposite to the
expected trend. However if one takes all the data into account then
the binary frequency increases from LCC to US to UCL. We stress here
again that the differences between the subgroups are not statistically
significant. An improved sample is needed to settle this issue.

It is interesting that the shift in the peak of the \veq\ distribution
correlates with the ages of the subgroups. This suggests that the
shift could also be due to evolutionary effects. Stars that evolve off
the main sequence expand their outer layer and this may lead to a
lowering of \veq. This effect should be most pronounced in the giants
and supergiants. The percentage of stars in these luminosity classes
is, however, very similar for the three subgroups. If one also takes
sub-giants into account then indeed UCL has the highest percentage of
evolved stars, but the difference with the other subgroups is
small. Moreover UCL and US have similar percentages of evolved
stars. Apart from the statistics, the magnitude of evolutionary
effects depends much on whether or not stars rotate differentially.

A drawback of the Lucy-rectification results presented above is that
the existence of a break-up velocity is not taken into account. Yet,
the \vsini\ values we derived for rapid rotators are based on the
value of the break-up velocity for the star in question. The
rectification technique presented by Bernacca (1970) does take the
existence of an upper limit on \veq\ into account. If there is a
break-up velocity then one can define an angle $\theta$ by:
$$
	\vsini=\vbreak\sin\theta\, .	\eqno\autnum
$$
It is clear that the inclination angle $i$ must then satisfy
$\theta\leq i\leq\pi/2$. So one has more information about the
rotation axes of the stars. If one assumes that the rotation axes are
randomly distributed in space with the restriction on $i$ given above,
the distribution of $i$, $f(i)$, can be derived from the observed
\vsini\ distribution:
$$
	f(i)=\sin i\int_0^{\sin i}
		{\phi(y)\over\sqrt{1-y^2}}\, dy\,.	\eqno\autnum
$$
The distribution of $\veq/\vbreak$ ($\eta$) is then given by:
$$
	f(\eta)={1\over\eta^2}\int_0^\eta
		{y^2\phi(y)\over\sqrt{\eta^2-y^2}\sqrt{1-y^2}}
			\, dy \,. 	\eqno\autnum
$$
For a derivation see Bernacca (1970). We have translated his formulas
to $y$ and $\eta$.

Thus, with the technique of Bernacca one can derive the distribution
of $i$ as well as $\eta$ and the distribution of $\eta$ is now given
by an integral which can be evaluated numerically. The results for the
three subgroups of Sco OB2 are shown in Fig.~13. We used the same
estimates of the observed PDF as for the Lucy-rectification as input
for Eq.~(18). The resulting distributions of $\eta$ (Fig.~13a) show
the same trend as the \veq\ distributions in Fig. 12. Namely, a shift
of the peak towards higher values of $\eta$ going from UCL to LCC to
US. Note that the up-turns in the distributions for LCC and UCL at
$\eta=1$ are probably not real but due to the singularity in the
integral in Eq.~(18). The distributions of $i$ (Fig.~13b) are
all very similar and essentially the difference from the distribution
given in Eq.~(1) is that there are less stars seen pole-on and more
seen equator-on.

\titlea{Effects of Rotation in the Hertzsprung-Russell Diagram}
Rapidly rotating stars have deformed surface layers and their surface
gravity varies from the pole to the equator. The atmospheric zones
around the pole have close to normal surface gravities and thus emit
an almost normal stellar spectrum. The zones close to the equator,
however, have lower surface gravities and also suffer from obscuration
due to gravity darkening. The stellar spectrum emitted from these
zones then corresponds to lower surface gravities as well as lower
effective temperatures. This implies that one expects changes in the
photometric colours of the stars with rotation. These changes are a
function of the inclination angle as well as the rotation speed.

These effects have been calculated by a number of authors for various
photometric systems (see e.g., CTC and references in Tassoul,
1978). In general stars are displaced away from the main-sequence and
the main-sequence for rotating stars always appears brighter than its
non-rotating counterpart. Stars observed pole-on are displaced
vertically towards higher luminosities and stars seen equator-on are
displaced more along the main sequence towards lower \teff.

\begfigside 6cm 12cm
\hskip -12.5cm
\psfig{figure=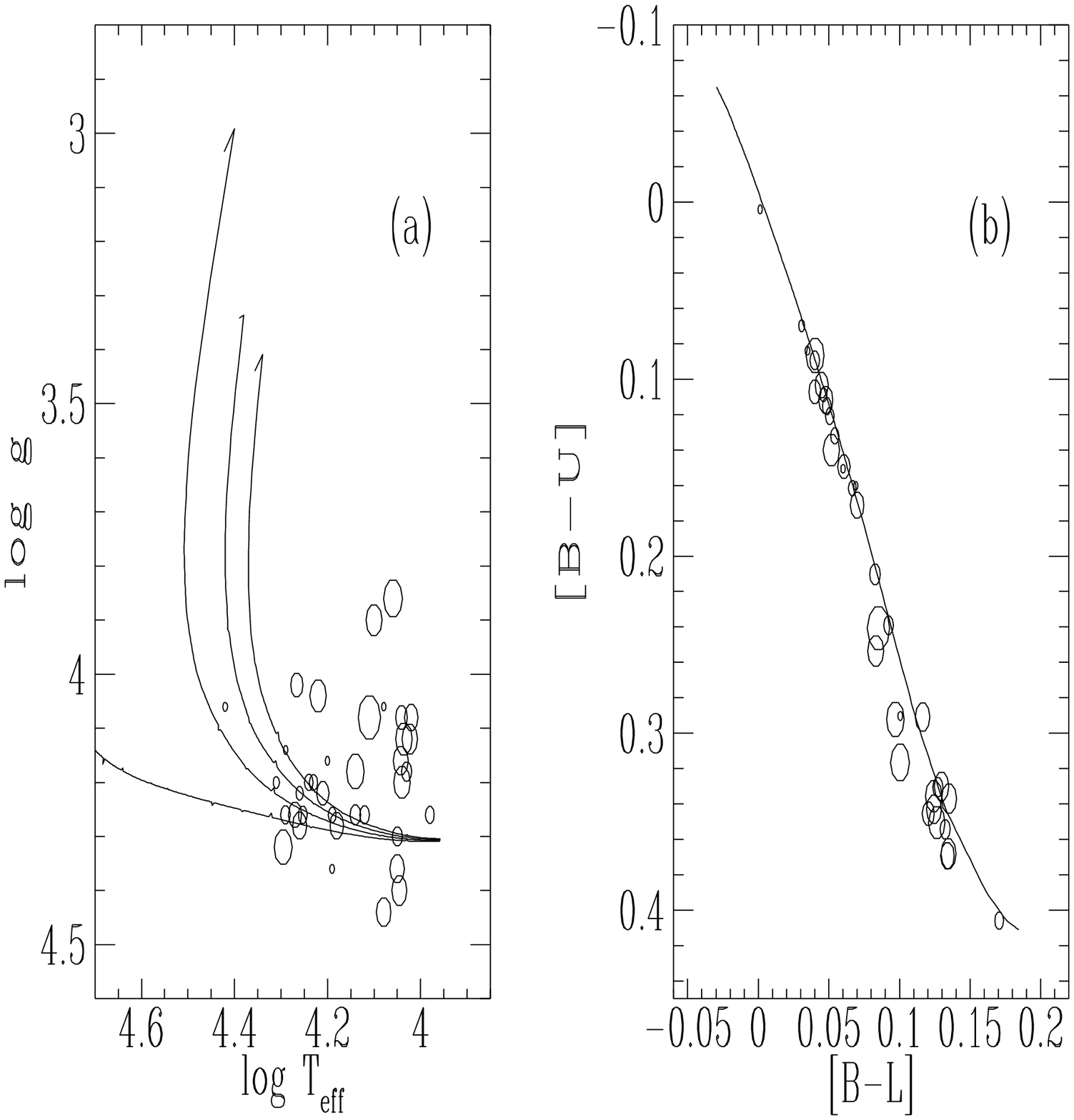,height=6cm,width=12cm}\hfil
\figure{14} {(a) The Hertzsprung-Russell diagram in
\logt\ and \logg\ for B1--B9 dwarfs that are members of Sco OB2 
(b) The Walraven \bminu\ vs.\ \bminl\ colour-colour diagram. In both
figures only single stars are included and the B9p stars are
excluded. The symbol size scales with $\vsini/\vbreak$ and isochrones
(from Schaller \ea 1992) are shown for 0, 5, 10 and $15\Myr$ in (a)
and for $0\Myr$ in (b).}
\endfig

The effects of rotation on the Walraven photometric system have not
been studied previously. In this section we investigate whether there
are stars for which the Walraven colours indeed correlate with
\vsini. We investigate unevolved members of the Sco OB2 association in
order to avoid the effects of stellar evolution on the colours. The
oldest subgroup in Sco OB2 has an age of 14--$15\Myr$ and this implies
that the upper mass limit for stars still on the main sequence is
about $12\msun$ (using the models by Schaller \ea 1992). This mass
corresponds to stars of spectral type $\sim {\rm B1}$ (see
e.g., Harmanec 1988). In the following we concentrate on the B1--B9
stars to find out whether there are effects of rotation on Walraven
colours.

For the stars in our data we used the Walraven photometry of de Geus
\ea\ (1990) and the value of \logt\ and \logg\ was taken from de Geus
\ea (1989). For some stars we had to derive \logt\ and \logg\ from the
photometry and that was done as described in de Geus \ea (1989) and
Brown \ea\ (1994). In Fig.~14a we show the HR-diagram in the
\logg-\logt\ plane for all B1--B9 main sequence dwarfs that are
members of Sco OB2. Binaries were excluded from this diagram as well
as B9p stars. The latter are known to have peculiar spectra and rotate
slowly as a class. In Fig.~14a the symbol size scales with
$\vsini/\vbreak$. It is clear that for $\logt<4.2$ the stars with
large values of $\vsini/\vbreak$ lie above the main sequence. In this
temperature range one finds B7--B9 stars. The Kelvin-Helmholtz
time-scale for these stars is $\sim 2$--$5\times10^5$~yr. Even for
members of US this is short enough for all B7--B9 stars to have
evolved onto the zero-age main-sequence. Thus we do not expect any
pre-main-sequence stars among this group.

The only other reason for these stars to lie above the main sequence
is that the transformation of the photometry to \logt\ and \logg\ is
in error. We therefore show the same sample of stars in the Walraven
\bminu\ vs.\ \bminl\ diagram in Fig.~14b. The ZAMS is
drawn in and the early type B-stars ($[B-L]<0.05$) all lie on or very
near this isochrone. In this colour-colour diagram the evolution of a
star away from the ZAMS is towards the lower left (see de Geus
\ea 1989). It is clear that for $[B-L]>0.05$ the stars are displaced
away from the ZAMS in the direction where evolved stars are
expected. If we calculate the distance of each B7--B9 dwarf from the
ZAMS (defined as the distance to the nearest point on the isochrone)
and correlate that distance with $\vsini/\vbreak$, we find a
correlation at the $95.2$\% confidence level. Thus the photometry is
indeed affected by rotation for the B7--B9 stars.

The largest decrease in \logg\ in Fig.~14a is $0.44$
for a star with $\vsini/\vbreak=0.84$. The latter value implies that
$w\geq0.96$. Judging from Fig.~2 in Collins (1963) the value of the
surface gravity then varies over a factor of $\sim 3$ from pole to
equator. This is consistent with our observed factor of $2.8$. We
conclude that the Walraven colours are affected by stellar rotation.

Note that some stars in Fig.~14b are located just above
the ZAMS. It concerns a B2V, two B4V, a B8V and two B9V stars. One of
the B4V stars is a star with emission lines. There is nothing peculiar
about the other stars. The positions of the normal B2V and B4V stars
can be explained as observational errors. The B8 and B9 stars however,
lie further away from the ZAMS than the errors allow. This can be
explained by the fact that for B7--B9 stars the empirical ZAMS for the
Walraven system, as given in Brand \& Wouterloot (1988), lies just
above the ZAMS defined by Schaller \ea (1992).

The lower surface gravity derived for the B7--B9 stars affects age
determinations based on isochrone fitting as well as the derivation of
mass distributions. The age of subgroups in the association will
appear higher because of the apparent presence of evolved B7--B9
stars. However, we note that the ages derived by de Geus \ea (1989)
are not affected by rotation. The stars in the B7--B9 range had little
weight in the isochrone fitting procedure. The same holds for the ages
derived for the Ori OB1 subgroups by Brown \ea (1994), who excluded
the B7--B9 stars from their isochrone fitting procedure. The IMFs
derived by de Geus (1992) and Brown \ea (1994) for Sco OB2 and Ori
OB1, respectively, are not affected by rotation. In both cases the
mass range used by the authors excluded B7--B9 stars.

\titlea{Conclusions and Future Work}
We have presented \vsini\ values for 156 established or probable
members of the Sco OB2 association. The projected rotational
velocities were derived from echelle spectra using three different
techniques, depending on the anticipated value of \vsini. We have
shown that the results for the different techniques are consistent and
that our results are consistent with the standard system of SCBWP. On
average the errors on our \vsini\ values are $\sim 10$\%.

We analyzed our own data together with data for members of Sco OB2
taken from the literature. The analysis of the \vsini\ distributions
for the members of Sco OB2 reveals that there are no significant
differences between the subgroups in Sco OB2. The binary stars in Sco
OB2 rotate slower on the whole than the single stars. This is
consistent with a picture in which angular momentum is transferred to
the binary orbit during star formation, and possibly further removed
due to tidal braking. For US we showed that the B7--B9 single stars
rotate significantly faster than the B0--B6 stars. This may be
consistent with the prediction about the distribution of rotational
velocities given by Mouschovias (1983). However, this result is derived
from inhomogeneous data.

Given the present data we cannot definitely exclude a form of random
distribution of intrinsic rotational velocities. The number of single
stars in the data is not enough to get statistically significant
results. Inclusion of the binaries complicates the picture due to the
different physical mechanism by which binary components may acquire a
low rotational velocity.

The rectification of the \vsini\ distributions by either the technique
of Lucy (1974, 1994) or that of Bernacca (1970), suggests that the peak
in the distribution of rotational velocities shifts to higher values
going from UCL to LCC to US. The interpretation of this finding in
terms of binary fraction per subgroup or evolutionary effects is not
possible with the present data.

For the Walraven photometric system we have now for the first time
shown conclusively that the colours of stars are affected by
rotation. This is most notably the case for the B7--B9 stars, where
the distance to the ZAMS in the \bminu\ vs.\ \bminl\ diagram
correlates with \vsini. However, we also conclude that the
determination of ages and mass distributions for Sco OB2 (de Geus
\ea 1989, de Geus 1992) and Ori OB1 (Brown \ea 1994) are not affected
by rotation.

The first step in improving the analysis of the \vsini\ distributions
in Sco OB2 is to get a complete sample in all subgroups. This entails
selecting members on the basis of kinematic data and performing a
complete census of the binary population. It is very timely that the
data from the HIPPARCOS astrometric mission will become available in
early 1996. The SPECTER consortium in Leiden (see de Zeeuw \ea 1994)
was granted observing time on this mission and we expect data for over
$10\,000$ stars of spectral type earlier than F8 in the direction of
nearby OB associations of which around 7000 are located in the
direction of Sco OB2. The data will be analyzed for membership. The
membership lists will greatly facilitate future extension of the
\vsini\ data as well as searches for binaries. In addition to proper
motions HIPPARCOS will also provide information on duplicity for
binaries with separations of more than $0.1$ arcsecond.

On the theoretical side it will be very useful to calculate the
expected effects of stellar rotation on Walraven colours. As a first
step the calculations of CTC can be used. We have used the Roche model
for rotating stars throughout this paper. This model is a very
simple approximation to reality and introduces a specific break-up
velocity. In the case of non-uniformly rotating stars it is not clear
that a break-up velocity exists and the observed equatorial velocity
depends very much on the internal distribution of angular
momentum. For a given mass a stronger degree of concentration of
angular momentum towards the centre of the star leads to a lower
observed \veq\ (see e.g., Bodenheimer 1971). So a consequence may be
that there are many more low \vsini\ stars than would be expected on
the basis of uniformly rotating models. Furthermore, the displacement
of stellar models in the HR-diagram changes and becomes more parallel
to the main-sequence if the central concentration of angular momentum
increases. In highly differentially rotating models the stellar
surface can develop a cusp at the poles and large variations in
surface gravity may result. The present measurements of \vsini\ still
depend on the Roche model for rotating stars (for the rapid rotators)
and on the correctness of approximating the atmosphere of a rotating
star as plane-parallel pieces of Kurucz atmospheres on the surface of
a rotating star. In order not to bias measurements of
\vsini\ better theoretical models are needed in which the atmospheric
structure follows from the parameters of the rotating star.

Finally, real constraints on the star formation process can only be
obtained if these data can be compared to models that make detailed
predictions of the distribution of rotational velocities, taking the
star formation process into account. The models should take into
account the early phases of star formation. In these phases the
details of the (possible) magnetic braking process may determine
whether a binary star is formed or not (e.g., Mouschovias
1991). Subsequently the angular momentum history of the protostellar
phase should be taken into account. Besides the angular momentum
history, the origin of stellar masses and their radii should also be
looked into, as these determine the observed value of
\vsini.

\acknow{ We thank H.\ Hensberge, R.S.\ Le Poole and P.T.\ de
Zeeuw for many fruitful discussions that greatly helped improve this
paper. M.\ David and E.J.\ de Geus are acknowledged for their help
in gathering and reducing the data. We thank R.J.\ Truax for kindly
providing the CTC models in digital form. M.~Perryman and J.~Lub are
acknowledged for comments that helped improve the final version of this
paper. This research was supported (in part) by the Netherlands
Foundation for Research in Astronomy (NFRA) with financial aid from
the Netherlands organization for scientific research (NWO). W.V.\
carried out this research in the framework of the project Service
Centres and Research Networks, initiated and financed by the Belgian
Federal Scientific Services (DWTC/SSTC).}

\begfig 8cm
\vskip -8cm
\psfig{figure=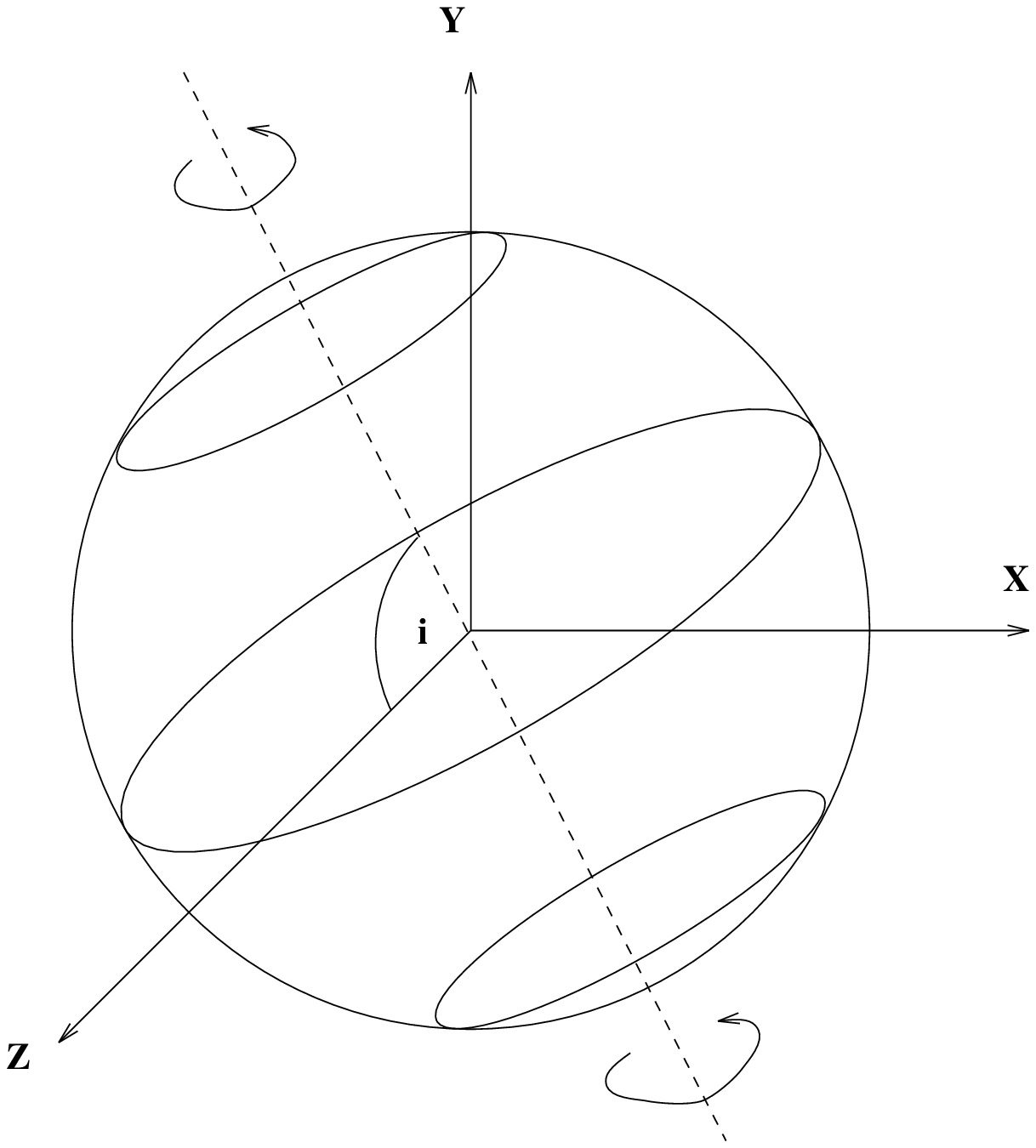,height=8cm}
\figure{15} {Frame of reference, attached to a rotating
star, that is to be used in deriving the rotational broadening
function. The Z-axis points towards the observer.}
\endfig

\appendix{A: The Rotational Broadening Function}
Following Tassoul (1978), we derive here the rotational broadening
function $A(x)$ for different limb-darkening laws. A rectangular frame
of reference is used with its origin at the centre of the star. The
$Z$-axis points towards the observer, and the $Y$-axis lies in the
plane passing through the axis of rotation and the line of sight. It
is assumed that the star is rotating uniformly, and that its observed
disc is a circle of radius $R$ (see also Fig.~15). For
convenience: $x=X/R$ and $y=Y/R$.

According to assumption C of the classical model for rotating stars
the line profiles in the spectrum of a star are constant over its
surface and are given by:
$$
I(x,y;\lambda-\lambda_{\rm c})=r(\lambda-\lambda_{\rm c})I_0(x,y)\,,
                     \eqno({\rm A}1)
$$
\noindent
where $I_0(x,y)$ is the continuum intensity emergent at the stellar
surface at position $(x,y)$ and $r(\lambda-\lambda_{\rm c})$ is the
ratio of the intensity in the spectral line to the continuum
intensity. The rotationally broadened spectral line is then defined
as:
$$
\bar r(\lambda-\lambda_{\rm c})=\int^{+1}_{-1}r(\lambda-\lambda_{\rm
c}-\lambda_{\rm c}x{\vsini\over c})A(x)\, dx\,.\eqno({\rm A}2)
$$


\noindent
Here $x\vsini=-v_{\scrscr z}$ and $A(x)$ is the rotational broadening
function given by:

$$
	A(x)={{\int_0^{(1-x^2)^{1/2}} I_0(x,y)\, dy}\over{
	\int_{-1}^{+1}\, dx\int_0^{(1-x^2)^{1/2}} I_0(x,y)\, dy}}\,. 
          \eqno({\rm A}3)
$$
\noindent
To calculate $A(x)$ a limb-darkening law has to be adopted. In Table 8
we list three forms for the limb-darkening law and the corresponding
rotational broadening functions. In the first column the
limb-darkening law is listed and in the second column the rotational
broadening function. In the limb-darkening laws
$\mu=\cos\theta=\sqrt{1-(x^2+y^2)}$, where $\theta$ is the angular
distance from the centre of the disc. The limb-darkening parameters
are $u$, $p$, $q$, $r$, and $s$.

\begtabfullwid
\tabcap{8} {Rotational broadening functions.}
\halign{\strut\hfil $\displaystyle{#}$ \quad\hfil
			  & \hfil $\displaystyle{#}$ \hfil \cr
\noalign{\smallskip\hrule\vskip4pt}
I_0(\mu)/I_0(1) & A(x) \cr
\noalign{\vskip4pt\hrule\medskip}
1-u(1-\mu) &
        {{2(1-u)(1-x^2)^{1/2}+
	{\pi\over 2}u(1-x^2)}\over{\pi(1-{u\over 3})}}\cr
\noalign{\bigskip\bigskip}
1-p(1-\mu)-q(1-\mu)^2 &
	{{2(1-p-q)(1-x^2)^{1/2}
	+{\pi\over 2}(2q+p)(1-x^2)-{4\over
	3}q(1-x^2)^{3/2}}\over{\pi(1-{{2p+q}\over 6})}}\cr
\noalign{\bigskip\bigskip}
1-r(1-\mu)-s(1-\sqrt\mu) &
	{{2(1-r-s)(1-x^2)^{1/2}+{\pi\over 2}r(1-x^2)+
	{3\over\sqrt{2\pi}}(\Gamma({1\over4}))^2s(1-x^2)^{3/4}}\over
	{\pi(1-r-s)+{{2\pi}\over3}r+{4\pi\over5}s}}\cr
\noalign{\medskip\hrule}}
\endtab

The first two limb-darkening laws are the well known linear and
quadratic approximation to limb-darkening and the third law is the
square-root approximation proposed by D\'\i az-Cordov\' es
\& Gim\' enez (1992).

If $S(\lambda)$ is the intrinsic spectrum, then the rotationally
broadened spectrum $S_{\rm r}(\lambda)$ is given by:
$$
S_{\rm r}(\lambda)=\int_{-1}^{+1}S\left(\lambda-\lambda x{\vsini\over c}\right)
	A(x)\, dx\,,\eqno({\rm A}4)
$$
or, substituting $z=\lambda x{\vsini\over c}$:
$$
S_{\rm r}(\lambda)=\int_{-\lambda_{\rm r}}^{+\lambda_{\rm r}}
	S(\lambda-z){1\over{\lambda_{\rm r}}}
	A\left({z\over{\lambda_{\rm r}}}\right)\, dz\,,\eqno({\rm A}5)
$$
where $\lambda_{\rm r}=\lambda {\vsini\over c}$.

\titleb{The Rotational Broadening Function in $\ln\lambda$-Space}
Rotational broadening of the spectrum depends on wavelength. This can
be seen in the convolution integral in Eq.~(A5) which is used
to calculate $S_{\rm r}(\lambda )$. The wavelength dependence is in
the sense that the rotational broadening function widens as $\lambda$
increases. This wavelength dependence is absent in $\ln\lambda$-space
(which is almost equivalent to pixel space for our spectra). By
substitution of $\lambda^\prime =\ln\lambda$ into the convolution
integral for $S_{\rm r}(\lambda )$ one finds:
$$
S_{\rm r}^\prime(\ln\lambda)=
	\int_{-\ln(1+{\vsini\over c})}^{-\ln(1-{\vsini\over c})}
	S^\prime(\ln\lambda-x^\prime)A^\prime(x^\prime)\, dx^\prime\,,
        \eqno({\rm A}6)
$$
where $S^\prime(\ln\lambda)$ is the spectrum and $A^\prime(x^\prime)$
the rotational broadening function in $\ln\lambda$-space.

The rotational broadening functions in $\ln\lambda$-space for the
different limb darkening laws can be found from the rotational
broadening functions in $\lambda$-space (listed above) as follows:
$$
\eqalignno{
A^\prime(x^\prime) & ={e^{-x^\prime}\over{\vsini /c}}
	A\left({{1-e^{-x^\prime}}\over{\vsini /c}}\right) & \cr
       &\quad -\ln\left(1+{\vsini\over c}\right)\leq x^\prime\leq
	-\ln\left(1-{\vsini\over c}\right)\,. & ({\rm A}7)\cr}
$$
This can be derived from the following:
$$
S_{\rm r}(e^{\lambda^\prime})=S^\prime_{\rm r}(\lambda^\prime)=
	\int_{-{e^{\lambda^\prime}\vsini\over
	c}}^{e^{\lambda^\prime}\vsini\over c}S(e^{\lambda^\prime}-z)
	{1\over {\lambda_{\rm r}}}A\left({z\over {\lambda_{\rm
	r}}}\right)\, dz\,.\eqno({\rm A}8)
$$
Setting $u=z/{e^{\lambda^\prime}}$, one finds:
$$
\eqalignno{
S^\prime_{\rm r}(\lambda^\prime)& =\int_{-{\vsini\over c}}^{{\vsini\over c}}
	S(e^{\lambda^\prime}(1-u))\times & \cr
        & \qquad\qquad {1\over{\vsini/c}}
	A\left({u\over{\vsini/c}}\right)\, du\,, & ({\rm A}9)\cr}
$$
since $S(y)=S^\prime(\ln y)$.
Substituting $x^\prime=-\ln (1-u)$ finally yields:
$$ 
\eqalignno{
S_{\rm r}^\prime(\lambda^\prime) & =
	\int_{-\ln(1+{\vsini\over c})}^{-\ln(1-{\vsini\over c})}
	S^\prime(\lambda^\prime-x^\prime)\times & \cr
	&\qquad\qquad {{e^{-x^\prime}}\over{\vsini/c}}
	A\left({{1-e^{-x^\prime}}\over{\vsini/c}}\right)\,
	dx^\prime\,. & ({\rm A}10)\cr}
$$

\titleb{The Width of Rotationally Broadened Lines}
At low values of the projected rotational velocity
($\vsini\la\allowbreak 80\kms$) the measured width of a line that is not
affected by other broadening mechanisms (Apart from thermal and
instrumental) can be directly converted into a value of \vsini. In
this paper we use metal lines for this purpose and measure their
FWHM by fitting Gaussians. This implies a measurement of the second
moment of the line profile. If the second moment of a Gaussian is
$\sigma$ then the FWHM is given by $2\sqrt{2\!\ln\!2}\sigma$. The
second moment of a broadened, intrinsically infinitely sharp line
centered on $\lambda_0$ is given by:
$$
\eqalignno{
\sigma^2(\lambda_0) & =\int_{\lambda_0-\lambda_{\rm
	r}}^{\lambda_0+\lambda_{\rm r}}
	(\lambda-\lambda_0)^2\bar r(\lambda)\, 
	d\lambda - & \cr
        &\qquad\qquad \left(\int_{\lambda_0-\lambda_{\rm
	r}}^{\lambda_0+\lambda_{\rm r}}(\lambda-\lambda_0)
	\bar r(\lambda)\, d\lambda\right)^2\,, & ({\rm A}11)\cr}
$$
with $\bar r(\lambda)$ in this case given by:
$$
\bar r(\lambda)=\int_{-\lambda_{\rm r}}^{+\lambda_{\rm r}}
\delta(\lambda-\lambda_0-z) {1\over\lambda_{\rm r}}
A\left({z\over\lambda_{\rm r}}\right)\, dz\,.\eqno({\rm A}12)
$$
Because of symmetry Eq.~(A11) reduces to:
$$
\sigma^2(\lambda_0)=\int_{-1}^{+1}\lambda_{\rm r}^2x^2A(x)\, dx\,.
    \eqno({\rm A}13)
$$
Note that the last integral contains an integrand in which only terms
of the form $x^2(1-x^2)^\nu$ occur, which are all symmetrical around zero. 
The Beta-function can be given as:
$$
\eqalignno{
B(x,y) & =2\int_0^1 t^{2x-1}(1-t^2)^{y-1}dt & \cr
       & \qquad\qquad\qquad ({\rm Re}(x)>0,\;{\rm Re}(y)>0)\,.
       & ({\rm A}14)\cr}
$$
So one can easily evaluate $\sigma^2$ using this expression for the
Beta-function and the relation between the Beta function and the
Gamma-function. For linear limb-darkening the result is:
$$
\sigma^2(\lambda_0)=\lambda_{\rm
r}^2\left({{3\over4}-{7\over20}u}\over{3-u}\right)\qquad\quad (\lambda_{\rm
r}=\lambda_0{\vsini\over c})\,.\eqno({\rm A}15)
$$
The results for quadratic and square-root limb-darkening are:
$$
\eqalignno{\sigma^2(\lambda_0) & =\lambda_{\rm r}^2 
         \left({{3\over2}+{1\over10}p-{6\over5}q}\over{6-2p-q}\right)\,,
	 & ({\rm A}16)\cr
	 \noalign{\vskip3pt}
         \sigma^2(\lambda_0) & =\lambda_{\rm r}^2
         \left({{{15\over4}-{7\over4}r-{13\over12}s}\over
	   {15-5r-3s}}\right)\,. & ({\rm A}17)\cr}
$$
The ECHELEC spectra that are being analyzed are recorded in
$\ln\lambda$-space and in that case one can show that the general
equation for the second moment $\sigma'$ is:
$$
\eqalignno{
(\sigma')^2 & =\int_{-1}^{+1}\ln^2(1-{\vsini\over c}y)A(y)\, dy - & \cr
            &\qquad\qquad \left(\int_{-1}^{+1}\ln(1-{\vsini\over c}y)A(y)\,
	    dy\right)^2\,. & ({\rm A}18)\cr}
$$
\noindent
The two integrals in the equation cannot be calculated
straightforwardly. However, in the limit $\vsini/c\ll 1$ the integrals
reduce to the same form as for $\sigma$ but without the dependence on
wavelength. Since the method is specifically applied in cases of low
values of \vsini\ the approximation should hold when determining
\vsini\ from the FWHM of the metal lines. For B3--B5 stars
($\teff\approx15\,000 {\rm K}$) the values of $u$, $p$, $q$, $r$, and
$s$ in the Str\"omgren $v$-band at $\logg=4.5$ are $0.408$, $0.159$,
$0.324$, $-0.102$, and $0.723$, respectively (taken from D\'\i
az-Cordov\'es \ea 1995). This leads to the following values for the
coefficients of $\lambda_{\rm r}$ in Eqs.~(A15), (A16) and (A17);
$0.234$, $0.210$, and $0.236$. This illustrates that the derived
values of \vsini\ are not very sensitive to the choice of
limb-darkening law.

\appendix{B: Fourier Filtering of Template and Target Spectrum}
As mentioned in Sect.~3.2 it is necessary to filter both the
target and the broadened template spectrum before comparing the two.
The filtering is done in Fourier-space and the high-pass filter is
constructed as follows. The information relevant to the rotation of a
star can be found in its spectrum on spatial scales $\sim2\vsini$
(in\kms). All components in the spectrum with longer spatial scales
than this value carry essentially no information on
\vsini\ and can be filtered out. So based on the highest value in the
\vsini\ search interval one can construct the high-pass filter in
Fourier-space. This filter is a step function with its edge tapered by
a narrow Gaussian. The edge of the filter is placed at the frequency
corresponding to the upper limit of the tested
\vsini\ values. In practice two extra frequencies below the edge value
are retained. This is to insure that one does not throw away too much
information. After filtering, the comparison between template and
target is done as described in Sect.~3.2.

The data do not contain a number of points equal to a power of two as
required by the FFT routine used. So in practice in order to avoid
creating large discontinuities in the data by zero padding we connect
the endpoints in the data by a straight line and subtract that line
from the data. The endpoints are often very noisy so we first discard
the outer 10 points at each end. Components with a spatial wavelength
smaller than $\sim260\kms$ were never filtered out in order to avoid
being dominated by small mismatches between template and target
spectrum. Signal to noise considerations also lead us to throw out 40
extra pixels at both edges of every spectral order (containing 500
pixels in total) before calculating the mean absolute difference
between template and target.

Note that before doing the comparison one has to make sure that the
template and target spectrum are shifted to the same radial
velocity. It turns out that aligning the spectra to within one pixel
is sufficiently accurate (see Appendix C). This can be done by eye.

\appendix{C: The Effects of Gravity- and Limb-Darkening
on Template Broadening} As discussed in Sect.~3.2 we expect
gravity- and limb-darkening to have more severe effects when one uses
template broadening to derive \vsini. We investigated this by
generating synthetic data from the ab initio models by CTC. We took
the models of the \ionI{He}4026\ line and the H$\gamma$ line and constructed
synthetic 'spectral orders' with the same characteristics as an
ECHELEC spectral order. The continuum was taken to be constant
(normalized flux 1) and Poisson noise was added artificially. The
artificial spectral order is calculated in $\ln\lambda$-space because
the models are given as a function of wavelength. Models with low
values of \vsini\ were used as templates.  This investigation was done
for B3 and B9 models in order to compare results for different
spectral types.

An example is shown in Fig.~16 for the \ionI{He}4026\ line. In
both cases the target spectrum is that of a star with $w=0.8$ and
$i=\deg{90}$ The corresponding values of \vsini\ are 250\kms and
213\kms, respectively, for the B3 and B9 target. The noise in the
target spectra corresponds to a signal to noise ratio of 250. This is
a typical value for order 142. The templates were non-rotating models
in both cases. The template broadening technique applied in these two
cases yields values for \vsini\ of 240\kms for the B3 target and
180\kms for the B9 target. The broadened template spectra in
Fig.~16 correspond to these values. Also plotted in
Fig.~16 are the same model spectra without the noise in order to
show the actual differences between the broadened template and the
target. In both cases one can see that the \ionI{He}4026\ line from
the ab initio model is narrower and less deep than the corresponding
classically broadened template. This effect is largest for the B9
star. In both cases \vsini\ is underestimated by template broadening,
but the error for the B3 star is less than 5\%. Note that for both
spectral types the fit of the noisy broadened template to the noisy
data looks good in the high-pass filter case even though the value of
\vsini\ is wrong.

For the parameters of the example shown ($S/N=250$, no radial velocity
shifts, zero-rotation template), template broadening gives the correct
results up to $w=0.8$ for B3 stars (\vsini\ is underestimated by
$\sim5\%$ for $w=0.8$, $i=\deg{90}$). For B3 stars rotating at
$w\ge0.9$ the value of \vsini\ can be underestimated up to 15\%. For
B9 stars template broadening only works well up to $w=0.5$ (\vsini\ is
underestimated by $\sim10\%$ for $w=0.5$, $i=\deg{90}$). For higher
values of $w$, \vsini\ is underestimated by at least 10--15\% and for
$w\ge 0.9$ the discrepancies between template broadening and model
fitting are so severe that depending on the inclination \vsini\ can be
overestimated as well as underestimated by $\sim20\%$.

\begfigwid 14cm
\vskip -14cm
\psfig{figure=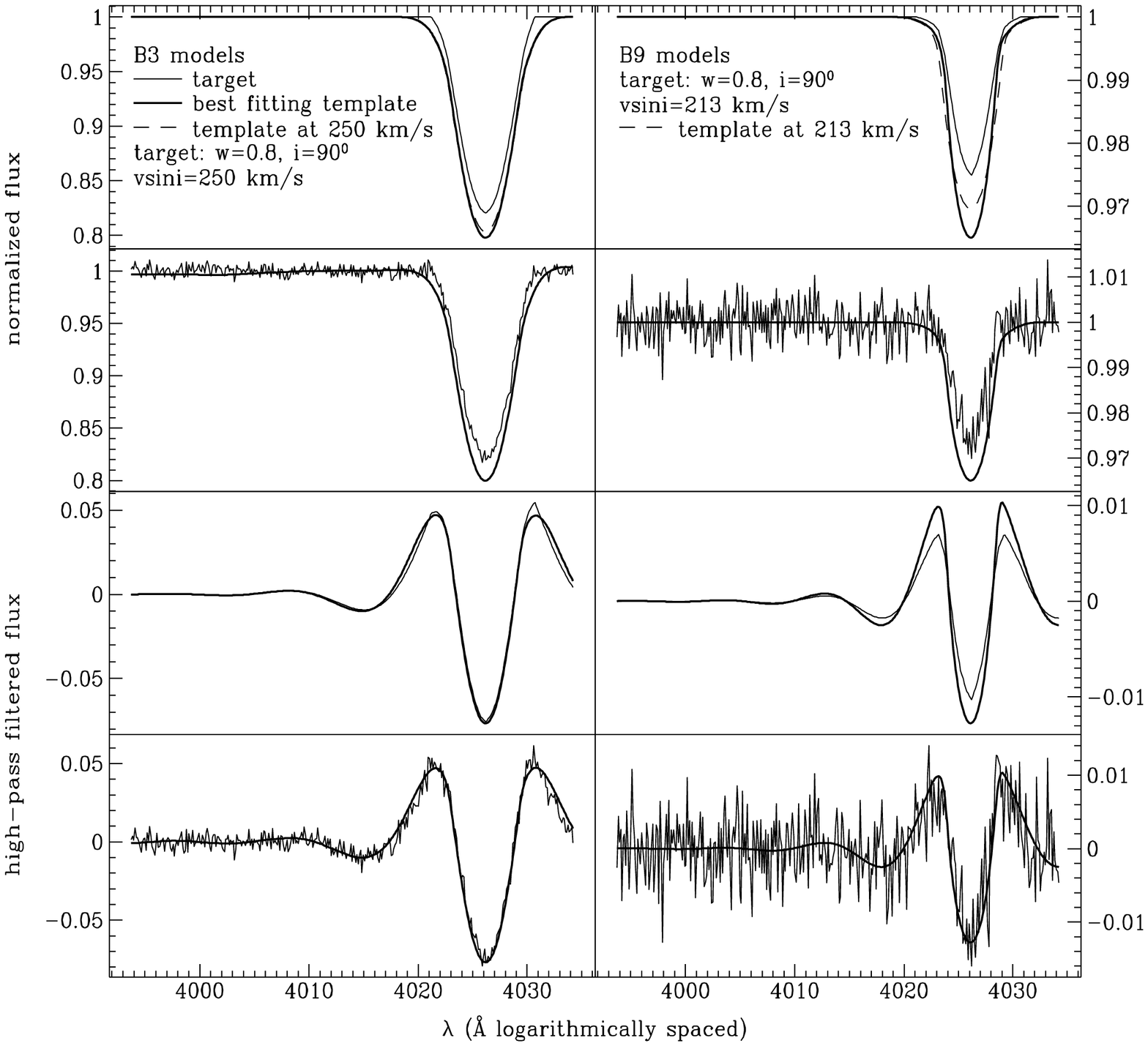,height=14cm,width=16cm}
\figure{16} {Example of template broadening applied to
synthetic data. The four panels on the left indicate the results for
B3 models. The panels on the right show the results for B9 models. The
top four panels show a direct comparison between the best fitting
broadened template (thick line) and target (thin line). The best fits
are based on the {\it filtered\/} data. In the top two panels the
dashed line is the broadened template corresponding to the real value
of \vsini\ of the object. The bottom panels show the comparison after
high-pass filtering of the data. The same results are shown for
synthetic data with and without noise. The noisy data have a signal to
noise ratio of 250. In both cases the template is a non-rotating star
and the target rotates at 80\% of the critical angular velocity.  The
inclination angle is \deg{90} in both cases. The resulting values of
\vsini\ are 250\kms and 213\kms, respectively, for the B3 and B9
targets. The values of \vsini\ found from template broadening are
240\kms and 180\kms, respectively. The broadened templates in this
figure correspond to the latter \vsini\ values. Note the differences
in vertical scale.}
\endfig

For the B3 models we investigated the effects of other parameters in
the synthetic data. At a lower signal-to-noise ratio in the spectrum
(S/N=100) the results are not affected. If one introduces small radial
velocity shifts of 1 to 2 pixels between template and object the
\vsini\ values from template broadening are off by 5--8\% at the
lowest intrinsic \vsini\ values. At higher values of \vsini\ the
results are not affected; it becomes less important at those values to
have the phases of template and object match exactly. If one uses a
template for which $\vsini\approx30\kms$ the value for the target is
underestimated even more. For $w\le0.8$ \vsini\ is now underestimated by
5--10\% and for higher values of $w$ the underestimates can be up to
17\%.

The results can be summarized by stating that the \vsini\ values for
template broadening of the \ionI{He}4026\ line can be trusted up to
$\sim200\kms$ for early type B3 stars. For the B9 stars one should
only trust template broadening for values of \vsini\ less than
120-150\kms.

The situation is reversed when using the H$\gamma$ line. Now it is the
B3 stars for which \vsini\ is underestimated at low values of $w$. The
value of \vsini\ from template broadening can be off by as much as
30\% at $w=0.5$. For B9 stars the value of \vsini\ from template
broadening can be trusted up to high values of $w$.

The difference in behaviour of \ionI{He}4026\ and H$\gamma$ with spectral
type is due to a difference in sensitivity to surface gravity. The
Balmer lines are most sensitive to surface gravity changes for the
early type B-stars. A reduction of the surface gravity causes the
wings of the Balmer lines to become less important whereas the line
strength in the core remains the same. This effect leads to
underestimated \vsini\ values and is strongest for the early-type
B-stars. The \ionI{He}4026\ line shows the opposite behaviour and it is the
late-type B-stars for which the line is most sensitive to changes in
surface gravity. In this case the whole line becomes weaker and
narrower.

The general conclusion is that one should be careful when using
template broadening results as soon as \vsini\ is larger than
$\sim120$--$150\kms$. Visual inspection of the results is required and
one should take the spectral type of the target star into account.

The role of limb-darkening was already discussed in
Sect.~3.1. In the simulations we also used the linear
limb-darkening law with parameter $u=0.4$ to calculate the broadened
template. The results suggest that the choice of limb-darkening law
does not affect the outcome by much. To investigate this a little
further we re-determined
\vsini\ for a couple of our stars with a different choice of
limb-darkening law. The two other forms we tried are the quadratic and
square-root approximations to limb-darkening (see e.g.,~D\'\i
az-Cordov\'es \& Gim\'enez 1992). We took a couple of stars and used a
limb-darkening law and parameters appropriate for the spectral type of
the stars. The results were not affected and we are confident that the
choice of limb-darkening law is not important when using template
broadening. The reason is that the differences due to limb-darkening
are obscured by the noise in the spectra.

\appendix{D: Notes to Table 2}

\runin{HD 94650:} The \ionI{He}4026\ line of this star is best fit with a B3
model. The appearance of the Helium lines in order 142 also points to
an earlier spectral type.

\runin{HD 95122:} The B5 model for \ionI{He}4026\ gives the best fit.

\runin{HD 100262:} This star is a supergiant and the measured \vsini\
may actually be determined by macro-turbulence in the atmosphere of
this star.

\runin{HD 101947:} The value of \vsini\ for this star was estimated by
eye from its spectrum. This value may also be a measure of
macro-turbulence rather than rotation. Note that this star is also
classified as G0Ia$^+$ (J.\ Lub, personal communication).

\runin{HD 104878:} For this star B9 models were used for \ionI{He}4026.

\runin{HD 105382:} The best template for the template broadening
method was HD 120709, which is a B5 star. Indeed, the star is
classified as B6IIIe in the Bright Star Catalogue.

\runin{HD 105580:} The best fitting template is HD 132955, a B3V
star. The spectrum of this star is clearly not of type B6. Houk and
Cowley (1975) classify this star as B3IV.

\runin{HD 107566:} The value of \vsini\ is taken from Uesugi \& Fukuda
(1981) and revised downwards by 5\% to make the value consistent with
the standard system given by SCBWP.

\runin{HD 107696:} The best fitting model for \ionI{He}4026\ is for spectral
type B5. The strength of the \ionI{He}4009\
line is clearly not consistent with spectral type
B9. Houk \& Cowley (1975) list B7Vn as spectral type for this star.

\runin{HD 108248:} HD 149438 (B0V, $\tau$ Sco) was used as template.

\runin{HD 108249:} This star is a double-lined spectroscopic
binary. The model fits were done by direct comparison (without Fourier
filtering) of the strongest component to the models. This makes this
\vsini\ value rather uncertain.

\runin{HD 108257:} The \ionI{He}4026\ line in this star is broader than the
models. The value of \vsini\ may thus be overestimated.

\runin{HD 108483:} The best fitting template is HD 96706, a B2V
star. The spectral type listed in Houk (1978) is B2V.

\runin{HD 109787:} The value of \vsini\ is taken from SCBWP.

\runin{HD 110432:} The best fit of the \ionI{He}4026\ line is for a B5 model.

\runin{HD 111123:} HD 149438 was used as template for method
2. HD 11123 is a non-radial pulsator and a $\beta$ Cephei variable.

\runin{HD 112078:} B3 models were used to fit the \ionI{He}4026\ line. The
line profile may be influenced by the presence of a secondary in the
spectrum.

\runin{HD 112091:} B4 models were used to fit the \ionI{He}4026\ line.

\runin{HD 112409:} B7 models were used to fit \ionI{He}4026.

\runin{HD 113703:} B3 models were used to fit \ionI{He}4026. Houk (1978)
lists spectral type B4V for this star.

\runin{HD 113791:} Listed as B3V by Houk (1978).

\runin{HD 113902:} B7 models were used to fit \ionI{He}4026.

\runin{HD 114529:} The value of \vsini\ for this star was estimated by
eye. It may be a double-lined spectroscopic binary.

\runin{HD 116072:} B3 models were used to fit \ionI{He}4026. The
spectral type of this star is B2.5Vn according to the Bright Star
Catalogue.

\runin{HD 116226:} B5 models were used to fit \ionI{He}4026. Houk (1978)
gives B5V for the spectral type of this star.

\runin{HD 118716:} B3III models were used for method 3. This star is a 
non-radial pulsator and a $\beta$ Cephei variable.

\runin{HD 118978:} B8 models were used to fit \ionI{He}4026.

\runin{HD 119361:} B7III models were used to fit \ionI{He}4026.

\runin{HD 120324:} This star is a non-radial pulsator.

\runin{HD 121190:} B7 models were used to fit \ionI{He}4026.

\runin{HD 121263:} The value of \vsini\ may be influenced by the
presence of a secondary component in the spectrum.

\runin{HD 124367:} B3 models were used to fit \ionI{He}4026.

\runin{HD 125238:} The stellar \ionI{He}4026\ line is much deeper than the
B3 model line. The value of \vsini\ may thus be underestimated.

\runin{HD 125745:} B6 models were used to fit \ionI{He}4026. The appearance of
the Helium lines in order 142 clearly suggests an earlier spectral
type.

\runin{HD 126341:} This star is a non-radial pulsator and a $\beta$
Cephei variable.

\runin{HD 126981:} B6 models were used to fit \ionI{He}4026. The appearance of
the Helium lines in order 142 suggests an earlier spectral type.

\runin{HD 127381:} HD 126341 (B2IV) was used as template.

\runin{HD 127972:} The value of \vsini\ is taken from SCBWP.

\runin{HD 128345:} B4 models were used to fit \ionI{He}4026. Houk (1978)
lists the spectral type of this star as B3/B4V

\runin{HD 129056:} This star is $\beta$ Cephei variable.

\runin{HD 131625:} B9 models were used to fit \ionI{He}4026. The models do
not fit well.

\runin{HD 132058:} HD 126341 was used as template. Houk (1978) lists
B2IV as spectral type for HD 132058.

\runin{HD 132851:} The value of \vsini\ was estimated from a visual
inspection of the spectrum.

\runin{HD 133937:} B3 models were used to fit \ionI{He}4026. The appearance of
the Helium lines in order 142 clearly suggests an earlier spectral
type. Houk (1978) lists the spectral type as B5/B7V.

\runin{HD 135876:} B6 models were used to fit \ionI{He}4026. The appearance of
the Helium lines in order 142 clearly suggests an earlier spectral
type. Houk (1978) lists the spectral type as B7V.

\runin{HD 136298:} B3 models were used to fit \ionI{He}4026. This star is a
$\beta$ Cephei variable.

\runin{HD 136664:} B3 models were used to fit \ionI{He}4026.

\runin{HD 136933:} The value of \vsini\ was estimated from a visual
inspection of the spectrum.

\runin{HD 137058:} B9 models were used to fit \ionI{He}4026. The models do
not fit well.

\runin{HD 137432:} HD 132955 (B3V) was used as template. The Bright
Star Catalogue lists the spectral type of this star as B4Vp.

\runin{HD 140784:} B4 models were used to fit \ionI{He}4026. The appearance of
the Helium lines in order 142 clearly suggests an earlier spectral
type. The Bright Star Catalogue lists the spectral type as B7Vn.

\runin{HD 141637:} B3 models were used to fit \ionI{He}4026. The spectrum
shows signs of a secondary component.

\runin{HD 142114:} The stellar \ionI{He}4026\ line is much deeper than the
B3 model line. The value of \vsini\ may thus be underestimated.

\runin{HD 142184:} The stellar \ionI{He}4026\ line is much deeper than the
B3 model line. The value of \vsini\ may thus be underestimated.

\runin{HD 142629:} The value of \vsini\ is taken from Uesugi \& Fukuda
(1981) and revised downwards by 5\% to make the value consistent with
the standard system given by SCBWP.

\runin{HD 142983:} The value of \vsini\ was estimated from a visual
inspection of the spectrum. According to SCBWP this star has a
projected rotational velocity of 400\kms. However, judging from the
spectrum this clearly is not the case. This star is also a non-radial
pulsator.

\runin{HD 143018:} The value of \vsini\ is taken from the Bright Star
Catalogue and the spectrum was inspected visually to check whether the
value listed is reasonable.

\runin{HD 143275:} B1 models were used to fit \ionI{He}4026\ and HD 149438
was used as template for method 2. The B1 model profiles are not
really deep enough to fit the data.

\runin{HD 144217:} HD 149438 was used as template.

\runin{HD 144294:} The stellar \ionI{He}4026\ line is much deeper than the
B3 model line. The value of \vsini\ may thus be underestimated.

\runin{HD 144987:} B7 models were used to fit \ionI{He}4026.

\runin{HD 145482:} The value of \vsini\ is influenced by the fact that
this star is a double-lined spectroscopic binary.

\runin{HD 147084:} The value of \vsini\ was estimated from a visual
inspection of the spectrum.

\runin{HD 147165:} This star is a $\beta$ Cephei variable.

\runin{HD 147628:} B6 models were used to fit \ionI{He}4026. The appearance of
the Helium lines in order 142 suggests an earlier spectral type.

\runin{HD 147933:} The value of \vsini\ may be influenced by the
presence of a secondary component in the spectrum.

\runin{HD 148184:} B4 models were used to fit \ionI{He}4026. This star shows
emission cores in the Balmer lines present in our spectra.

\runin{HD 149757:} The value of \vsini\ is taken from SCBWP. This star
is a non-radial pulsator.

\runin{HD 151804:} The value of \vsini\ was estimated from a visual
inspection of the spectrum.

\runin{HD 151890:} B1 models were used to fit \ionI{He}4026. The value of
\vsini\ may be influenced by the secondary component in the spectrum.

\runin{HD 154204:} B4 models were used to fit \ionI{He}4026. The appearance of
the Helium lines in order 142 suggests an earlier spectral type. The
Bright Star Catalogue lists the spectral type as B6IV.

\runin{HD 157042:} The value of \vsini\ was estimated from a visual
inspection of the spectrum. This star is a non-radial pulsator.

\runin{HD 157056:} This star is a $\beta$ Cephei variable.

\runin{HD 157246:} The value of \vsini\ is taken from SCBWP. This star is a 
non-radial pulsator.

\runin{HD 158427:} The value of \vsini\ may be influenced by the
presence of a secondary component in the spectrum.

\runin{HD 158926:} The value of \vsini\ was derived from the spectrum
by visual inspection. The \vsini\ value listed in SCBWP is
190\kms. This star is a $\beta$ Cephei variable.

\runin{HD 168905:} The stellar \ionI{He}4026\ line is much deeper than the
B3 model line. The value of \vsini\ may thus be underestimated.

\runin{HD 171034:} HD 126341 was used as template.

\begref{References}

\ref
Bernacca, P.L., 1970, in Stellar Rotation, ed.\ A.\ Slettebak,
Reidel Dordrecht, p.\ 227

\ref
Bertiau, F.C., 1958, \apj 128, 533

\ref
Blaauw, A., 1946, PhD.\ Thesis, Groningen University 

\ref
Blaauw, A., 1964, \araa 2, 213

\ref
Blaauw, A., 1978, in Problems of Physics and Evolution of the Universe, ed.\
L.\ Mirzoyan, Yerevan, USSR, p.\ 101

\ref
Blaauw, A., 1991, in The Physics of Star Formation and Early Stellar
Evolution, eds.\ C.J. Lada \& N.D.\ Kylafis, NATO ASI Series C,
Vol.\ 342, p.\ 125

\ref
Bodenheimer, P., 1971, \apj167, 153

\ref
Bodenheimer, P., Ruzmaikina, T., Mathieu, R.D., 1993, in Protostars
and Planets III, eds.\ E.H.\ Levy \& J.I.\ Lunine, The University of
Arizona Press, p.\ 367

\ref
Brand, J., Wouterloot, J.G.A., 1988, \aapss 75, 117

\ref
Brown, A.G.A., de Geus, E.J., de Zeeuw, P.T., 1994, \aap 289, 101

\ref
Chandrasekhar, S., M\"unch, G., 1950, \apj 111, 142

\ref
Collins, G.W., II, 1963, \apj 138, 1134

\ref
Collins, G.W., II, Sonneborn, G.H, 1977, \apjss 34, 41

\ref
Collins, G.W., II, Truax, R.J., 1995, \apj 439, 860

\ref
Collins, G.W., II, Truax, R.J., Cranmer, S.R., 1991, \apjss 77, 541 (CTC)

\ref
Day, R.W., Warner, B., 1975, \mn 173, 419

\ref
Deutsch, A.J., 1970 in Stellar Rotation, ed.\ A.\ Slettebak,
Reidel Dordrecht, p.\ 207

\ref
D\'\i az-Cordov\' es, J., Gim\' enez, A., 1992, \aap 259, 227

\ref
D\'\i az-Cordov\' es, J., Claret, A., Gim\' enez, A., 1995, \aapss
110, 329

\ref 
de Geus, E.J., 1992, \aap 262, 258

\ref
de Geus, E.J., de Zeeuw, P.T., Lub, J., 1989, \aap 216, 44

\ref
de Geus, E.J., Lub, J., van de Grift, E., 1990, \aapss 85, 915

\ref
Gray, D.F., 1992, The Observation and Analysis of Stellar
Photospheres, Cambridge Astrophysics Series

\ref
Harmanec, P., 1988, Bull.\ Astron.\ Inst.\ Czechosl., 39, 329

\ref
Hensberge, H., van Dessel, E.L., Burger, M., \ea\ 1990, The Messenger,
61, 20

\ref
Hoffleit, D., Jaschek, C., 1982, The Bright Star Catalogue ($4^{\rm th}$
revised edition), Yale University Observatory

\ref
Hoffleit, D., Saladyga, M., Wlasuk, P., 1983, A Supplement to the
Bright Star Catalogue, Yale University Observatory

\ref
Houk, N., Cowley, A.P., 1975, University of Michigan Catalogue of
Two-Dimensional Spectral Types for the HD Stars: Volume 1, University
of Michigan, Ann Arbor

\ref
Houk, N., 1978, University of Michigan Catalogue of Two-Dimensional
Spectral Types for the HD Stars: Volume 2, University of Michigan, Ann
Arbor

\ref
Kurucz, R.L., 1991, Harvard Preprint 3348

\ref
Lamers, H.J.G.L.M., Leitherer, C., 1993, \apj 412, 771

\ref
Levato, H., Malaroda, S., Morrell, N., Solivella, G., 1987, \apjss 64,
487

\ref
Lucy, L.B., 1974, \aj 79, 745

\ref
Lucy, L.B., 1994, \aap 289, 983

\ref
Maeder, A., 1971, \aap 10, 354

\ref
Mouschovias, T.Ch., 1983, in Solar and Stellar Magnetic Fields:
Origins and Coronal Effects, IAU Symp.\ 102, ed.\ J.O.\ Stenflo,
Dordrecht, Reidel, p.\ 479

\ref
Mouschovias, T.Ch., 1991, in The Physics of Star Formation and Early
Stellar Evolution, eds.\ C.J. Lada \& N.D.\ Kylafis, NATO ASI Series
C, Vol.\ 342, p.\ 61

\ref
Press, W.H., Teukolsky, S.A., Vetterling, W.T., Flannery, B.P., 1992,
Numerical Recipes in Fortran: The Art of Scientific Computing, Second
Edition, Cambridge University Press

\ref
Rajamohan, R., 1976, Pramana, Vol.\ 7, 160

\ref 
Schaller, G., Schaerer, D., Meynet, G., Maeder, A., 1992,
\aapss, 96, 269

\ref
Siegel, S., Castellan, N.J., Jr., 1988, Nonparametric Statistics for
the Behavioral Sciences, McGraw-Hill Singapore

\ref
Slettebak, A., 1954, \apj 119, 146

\ref
Slettebak, A., 1955, \apj 121, 653

\ref
Slettebak, A., 1956, \apj 124, 173

\ref
Slettebak, A., 1966a, \apj 145, 121

\ref
Slettebak, A., 1966b, \apj 145, 126

\ref
Slettebak, A., 1968, \apj 151, 1043

\ref
Slettebak, A., Howard, R.F., 1955, \apj 121, 102

\ref
Slettebak, A., Collins, G.W., II, Boyce, P.B., White, N.M., Parkinson,
T.D., 1975, \apjss 29, 137 (SCBWP)

\ref
Spitzer, L., Jr., 1968, Diffuse Matter in Space, Interscience Tracts on
Physics and Astronomy

\ref 
Strai\v zys, V., Kuriliene, G., 1981, \apss 80, 353

\ref
Tassoul, J.-L., 1978, Theory of Rotating Stars, Princeton Series in
Astrophysics

\ref
Turon, C., Cr\'ez\'e, M., Egret, D., G\'omez, A., \ea, 1992, ESA SP-1136

\ref
Uesugi, A., Fukuda, I., 1981, Revised Catalogue of Stellar Rotational
Velocities, France: CDS Strasbourg

\ref
Verschueren, W., Brown, A.G.A., Hensberge, H., David, M., Le Poole,
R.S., de Geus, E.J., de Zeeuw, P.T., 1996, submitted to \pasp{}

\ref
Verschueren, W., \ea, 1996, in preparation

\ref
Vio, R., Fasano, G., Lazzarin, M., Lessi, O., 1994, \aap 289, 640

\ref
Wolff, S.C., Edwards, S., Preston, G.W., 1982, \apj 252, 322

\ref
de Zeeuw, P.T., Brown, A.G.A., Verschueren, W., 1994, in Galactic and
Solar System Optical Astrometry: Observation and Application, eds.\
L.V.\ Morrison \& G.F.\ Gilmore, Cambridge University Press, p.\ 215

\endref

\bye